\documentclass[fleqn,usenatbib]{mnras}

\usepackage{newtxtext,newtxmath}

\usepackage[T1]{fontenc}
\usepackage{ae,aecompl}
\usepackage{hyperref}

\usepackage{graphicx}	
\usepackage{amsmath}	
\usepackage{amssymb}	

\usepackage{booktabs, siunitx, caption}

\newcommand{\totalnumber}{82 }
\newcommand{\totalnumberminusone}{81 }
\newcommand{\quadrupolemodenumber}{17 }
\newcommand{\totalnumberminusfour}{78 }
\newcommand{\excludednumberHRdiagram}{4 }

\title[Rossby modes in \totalnumber $\gamma$\,Dor stars]{Period spacings of $\gamma$ Doradus pulsators in the \textit{Kepler} field: Rossby and gravity modes in \totalnumber stars}

\author[G. Li et al.]{
Gang Li$^{1,2\thanks{E-mail: gali8292@uni.sydney.edu.au}}$,
Timothy Van Reeth$^{1,2}$,
Timothy R. Bedding$^{1,2\thanks{E-mail: tim.bedding@sydney.edu.au}}$,
Simon J. Murphy$^{1,2}$,
\newauthor{Victoria Antoci$^2$}
\\
$^1$Sydney Institute for Astronomy (SIfA), School of Physics, University of Sydney 2006, Australia\\
$^2$Stellar Astrophysics Centre, Department of Physics and Astronomy, Aarhus University, Ny Munkegade 120, DK-8000 Aarhus C, Denmark
}

\date{Accepted XXX. Received YYY; in original form ZZZ}

\pubyear{2018}

\usepackage{etoolbox}
\makeatletter
\patchcmd\@combinedblfloats{\box\@outputbox}{\unvbox\@outputbox}{}{%
   \errmessage{\noexpand\@combinedblfloats could not be patched}%
}%
 \makeatother

\begin{document}
\label{firstpage}
\pagerange{\pageref{firstpage}--\pageref{lastpage}}
\maketitle

\begin{abstract}
Rossby modes are the oscillations in a rotating fluid, whose restoring force is the Coriolis force. They provide an additional diagnostic to understand the rotation of stars, which complicates asteroseismic modelling. We report \totalnumber $\gamma$\,Doradus stars for which clear period spacing patterns of both gravity and Rossby modes have been detected. The period spacings of both show a quasi-linear relation with the pulsation period but the slope is negative for the gravity modes and positive for the Rossby modes. Most Rossby modes have $k=-2, m=-1$. For only one star a series of $k=-1,m=-1$ modes is seen. For each pattern, the mean pulsation period, the mean period spacing, and the slope are measured. We find that the slope correlates with the mean period for Rossby mode patterns. The traditional approximation of rotation is used to measure the near-core rotation rate, assuming the star rotates rigidly. We report the near-core rotation rates, the asymptotic period spacings, and the radial orders of excited modes of these \totalnumber main-sequence stars. The near-core rotation rates lie between $0.6\,\mathrm{d^{-1}}$ and $2.3\,\mathrm{d^{-1}}$. Six stars show surface rotation modulations, among which only KIC\,3341457 shows differential rotation while the other five stars have uniform rotations. The radial orders of excited modes show different distributions for the dipole and quadrupole gravity modes versus the Rossby modes. 
\end{abstract}




\section{Introduction}

\textcolor{black}{Rotation is one of the key parameters that decides how stars evolve, since it affects many physical processes of the stellar interior \citep[e.g.][]{Maeder_2009, Mathis_2013}. For example, differential rotation induces an extra shear mixing and changes the local abundances of chemical elements. It can transport fuel into the nuclear burning regions and extend the lifetime of the star \citep[e.g.][]{VanReeth_2018, Prat_2018}. Differential rotation, both in radial and latitudinal directions, has been seen in the Sun and some sun-like stars \citep[e.g.][]{Thompson_1996, Couvidat_2003, Benomar_2018}, and A-type main sequence stars \citep[e.g.][]{Hatta_2019}. However, our understanding of rotation is still incomplete \citep[e.g.][]{Aerts_2014, Cazorla_2017}. }

In the rotating frame, the Coriolis force is defined to maintain Newton's laws of motion. This kind of inertial force acts as the restoring force for Rossby modes \citep[r-modes,][]{Papaloizou_1978_first_r_mode}. Rossby modes are inertial modes, unlike the well-known gravity modes and pressure modes from asteroseismology, whose restoring forces are buoyancy and the pressure gradient, respectively. 
Rossby modes involve global toroidal motions coupling with the spheroidal motion by the Coriolis force, whose compression and expansion lead to the temperature perturbations and hence make the Rossby modes visible \citep{Pedlosky_1982_book, Saio_2018_Rossby_mode}. This wave propagates in the direction retrograde to the rotation and is confined to the mid-latitudes \citep{Saio_1982_Rossby_mode, Lee_1997_rotating}, introducing the possibility of measuring the stellar inclination in future \citep{Saio_2018_Rossby_mode}. Rossby waves also have discrete frequencies that are smaller than the rotation frequency in the inertial frame \citep{Provost_1981_eigenfreq_of_Rossby_mode}. We refer the interested reader to \cite{Saio_2018_Rossby_mode}, and the references therein, for a more extensive theoretical discussion about r-modes.

Rossby waves are seen or predicted in many types of rotating systems. They appear in planetary atmospheres \citep{Rossby_1939} and play a key role in accurate weather forecasting \citep[e.g.][]{Screen_2014_weather_extremes}. The Sun rotates slowly but the Rossby waves still can be observed by tracking the coronal brightpoints and the south-north helioseismic travel times \citep{McIntosh_2017_sun_rossby_mode, loptien_2018_sun_rossby_mode, Liang_2018}. They have the potential to improve our understanding of energy transfer, solar activity, and space weather \citep{Zaqarashvili_2015}. Many questions remain. For example, how do the Rossby waves affect the protoplanetary disc \citep{Lovelace_2014} and how do they affect the evolution of DA white dwarfs and neutron stars \citep{Saio_1982_Rossby_mode, Berthomieu_1983, Andersson_1998, Brown_2000}?

In this paper, we conduct an observational study of $\gamma$\,Doradus stars with resolved gravity and Rossby modes. The brightness variations of $\gamma$\,Dor stars have been observed for several decades \citep[e.g.][]{Cousins_1963_first_gdor,Balona_1994}. They are F to A-type main sequence stars with typical masses between $1.4\,\mathrm{M_\odot}$ and $2.0\,\mathrm{M_\odot}$, burning hydrogen in their convective cores \citep{VanReeth_2015_detection_method}. 
The pulsations in $\gamma$\,Dor stars are mainly high radial order ($20 \lesssim n \lesssim 120$) low degree ($l\leq 4$) gravity modes \citep[e.g.][]{Balona_1994, Kaye_1999, VanReeth_2016_TAR, Saio_2018}, which are useful for investigating the interior of stars since they have the highest mode energy in the near-core regions \citep{Triana_2015, VanReeth_2016_TAR}. \cite{Shibahashi1979} pointed out that the periods of high radial order g-modes ($n\gg l$) are equally spaced for chemically homogeneous non-rotating stars. \cite{Miglio_2008} considered the effect of chemical composition gradients and found that dips or glitches appear in the period spacing patterns. 
Rotation also affects period spacings, causing the spacing of the prograde ($m>0$) and zonal ($m=0$) modes to decrease linearly with increasing period, and overall spacing to increase for the retrograde ($m<0$) modes in the inertial reference frame \citep{Bouabid_2013, Ouazzani_2017}.

The long pulsation periods ($\sim$1\,d), small period spacings ($\sim$1000\,s) and the low pulsation amplitudes ($\sim 0.1\%$ variation) are challenging for ground-based observations. Thanks to the high precision and the long time-base of \textit{Kepler} data \citep{Borucki_2010,Koch_2010}, period spacing patterns of tens of $\gamma$\,Dor stars have been detected. Some of these stars rotate slowly, hence nearly regular period spacings and rotation splittings are seen \citep[e.g.][]{Kurtz_2014, Saio_2015, Keen_2015,Murphy_2016, Li_2018}. However, most $\gamma$\,Dor stars rotate rapidly and the period spacings change significantly with period \citep[e.g.][]{Bedding_2015, VanReeth_2015, Saio_2018, Christophe_2018}. Besides g-modes, r-modes were also found in $\gamma$\,Dor stars by \cite{VanReeth_2016_TAR} and \cite{Saio_2018_Rossby_mode}. The resolved r-modes generally show a quasi-linearly increasing period spacing with increasing period. However, the period spacing in r-modes drops at the larger period region \citep[see the observation and theory in][]{Saio_2018_Rossby_mode}. \cite{Saio_2018_Rossby_mode} also explained observed power excesses in the periodogram as unresolved Rossby modes, in which the discrete frequencies are below the frequency resolution.

Here, we report \totalnumber $\gamma$\,Dor stars in which the period spacing patterns of g-modes and r-modes are both seen. The co-existence of prograde and retrograde modes spacings breaks the correlation between the asymptotic spacing and the near-core rotation rate, and allows us to determine their values more precisely \citep{VanReeth_2016_TAR}. 
The resolved r- and g-mode spacings also give many new possibilities for future research. They can be used in detailed asteroseismic modelling analyses \citep[e.g.][]{Aerts_2018_ApJS} to study the inner physics of main-sequence stars, such as differential rotation and angular momentum transport \citep{Aerts_2018, VanReeth_2018}.

In Section~\ref{sec:Data reduction}, we briefly review the light curve reduction and pattern identification introduced by \cite{Li_2018}. In Section~\ref{sec:near core rotation}, we show how the traditional approximation of rotation (TAR) is used to model the observed patterns and reveals the near-core rotation rate. In Section~\ref{sec:observational_results}, we report the observational results, such as the HR diagrams of our stars using \textit{Gaia} data, the slope-period relations of g- and r-modes, near-core and surface rotation rates, and radial order distributions. 
Finally, we summarise our findings in Section~\ref{sec:conclusions}. 

\section{Pattern extraction}\label{sec:Data reduction}
We used 4-year \textit{Kepler} long-cadence (LC; 29.45-min sampling) light curves from the multi-scale MAP data pipeline \citep{Stumpe_2014}. In each quarter, the light curve was divided by a second-order polynomial fit to remove any slow trend. We computed the Fourier transform and extracted the frequencies until the signal to noise ratio (S/N) was smaller than 3. 
The frequency uncertainty is given by 
\begin{equation}
\sigma\left(f\right)=0.44\frac{\langle a \rangle }{a}\frac{1}{T},\label{equ:freq_uncertainty}
\end{equation}
where $\langle a \rangle$ is the noise level in the amplitude spectrum, $a$ is the amplitude of peak, and $T$ is the total time span of the \textit{Kepler} data $(T\simeq 1470\,\mathrm{d})$ \citep{Montgomery1999, Kjeldsen2003}. Considering the S/N threshold, the frequency uncertainty in our work is smaller than $1\times 10^{-4}\,\mathrm{c/d}$. 

Combination frequencies are seen in some $\gamma$\,Dor stars. They form several frequency groups in the power spectrum but the understanding of them is still an open question \citep{Kurtz2015, Saio_2018}. Likely combination frequencies were determined by the condition
\begin{equation}
|n_i f_i+n_j f_j-f_k|<0.0002\,\mathrm{c/d},
\end{equation}
where $f_i$ and $f_j$ are parent frequencies, $n_i$ and $n_j$ are the coefficients, and $f_k$ is the combination candidate. We selected the 20 highest peaks as parent frequencies and set $|n_i|+|n_j|\leq2$. In our work, the patterns dominated by the combination frequencies could be explained as the higher-degree prograde sectoral oscillations following \cite{Saio_2018}.

A cross-correlation algorithm was implemented to detect the period spacing patterns between 0.2\,d and 2\,d \citep{Li_2018}. 
\textcolor{black}{We built a template to imitate the shape of periodogram, which was specified by the central peak's period $P_\mathrm{c}$, the central peak's period spacing $\Delta P_\mathrm{c}$, the slope $\Sigma \equiv \mathrm{d}\Delta P/ \mathrm{d} P$, and the number of peaks. The product of the template and the observed periodogram was used to measure the goodness of fit. The larger the product, the better the template described the observed pattern. }

Then, the pattern is fitted by the formula
\begin{equation}
P_i=\Delta P_0 \frac{\left(1+\Sigma\right)^i-1}{\Sigma}+P_0=\Delta P_0\left(n'+\epsilon \right)\label{equ:P_i}
\end{equation}
with the assumption that the period spacing changes linearly with period. Here, $P_i$ is the $i^\mathrm{th}$ peak, $P_0$ is the first peak, $\Delta P_0$ is the first period spacing, $\Sigma$ is the slope in the linear assumption, $n'\equiv \frac{\left(1+\Sigma\right)^i-1}{\Sigma}$ is the normalised index, and $\epsilon$ is the ratio $P_0/\Delta P_0$ \citep{Li_2018}. Due to the occasional missing peaks, not all the period spacings can be calculated. This formula can be used to include isolated peaks, where there is no adjacent peak to calculate the period spacing value, and is helpful in calculating the radial order differences. \textcolor{black}{Equation~\ref{equ:P_i} also allows us to make an \'{e}chelle diagram for less clear period spacing patterns and guide the pattern identification. }

We can assign an angular degree $l$ and an azimuthal order $m$ to gravity modes. For the Rossby modes, the angular degree $l$ is not defined and the value $k$ is used instead \citep{Lee_1997_rotating}. For gravity modes, $k=l-|m|$. This paper adopts the convention that positive $m$ represents the prograde modes while negative $m$ represents retrograde modes.

\subsection{An example: KIC\,3240967}

\begin{figure*}
\centering
\includegraphics[width=1\linewidth]{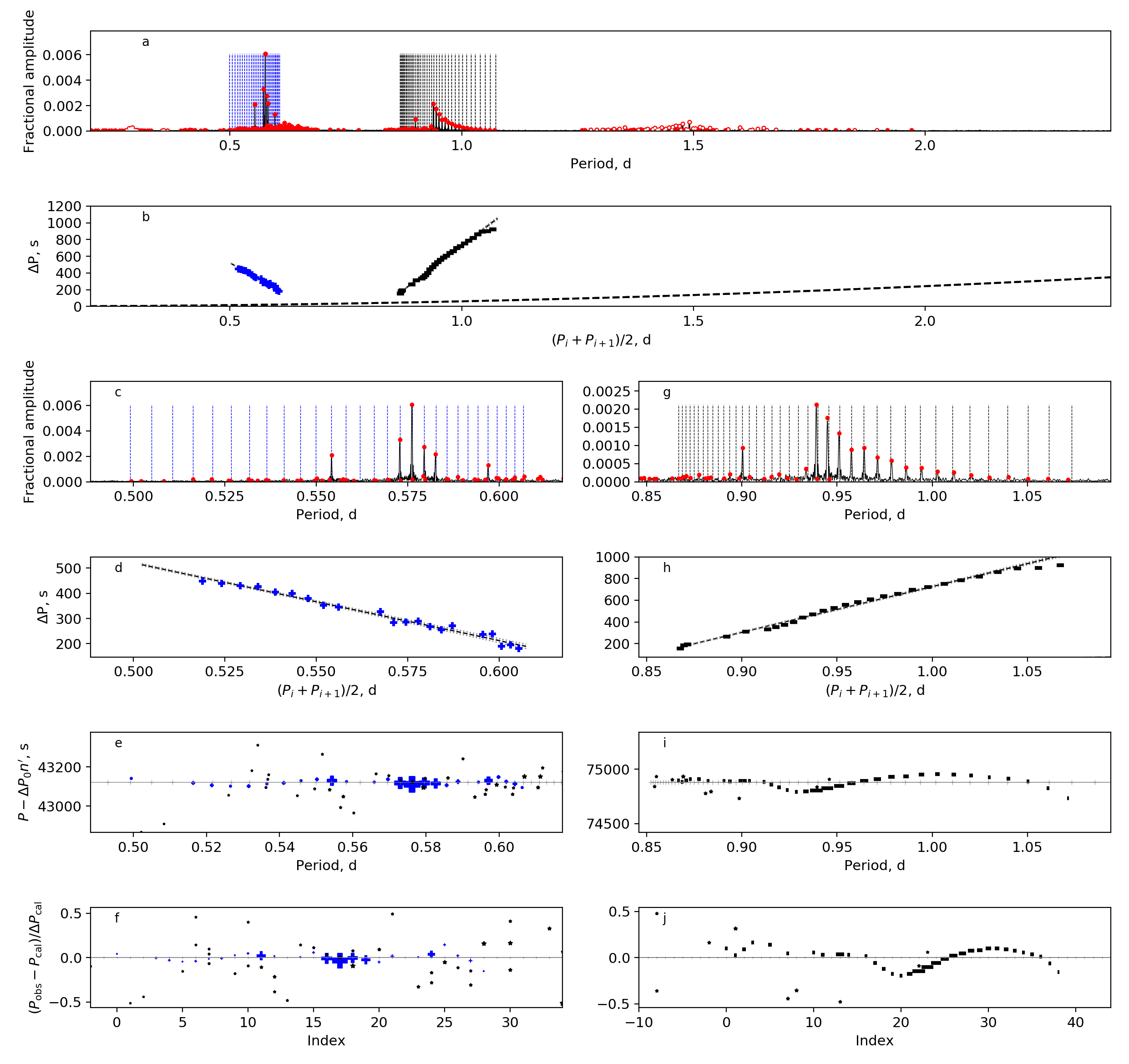}
\caption{The g and r-mode patterns of KIC\,3240967. Panel a: the amplitude spectrum with x-axis of period. The solid red circles present the detected independent frequencies while the open red circles show the combination frequencies. The vertical dashed lines are the linear fits for each pattern. The x-axis range is set from 0.2\,d to 2.4\,d for consistency for all stars. We found two independent frequency groups around 0.55\,d and 0.95\,d and two combination frequency groups around 0.3\,d and 1.5\,d. There are two period spacing patterns. The blue one on the left is the $l=1, m=1$ g-modes while the black one on the right is the $k=-2, m=-1$ r-modes. Panel b: the period spacing patterns of KIC\,3240967. The linear fits and uncertainties are shown by the black and grey dashed lines. The blue plus symbols are the g-modes and the black minus symbols are the r-modes. Panels c and d: the detail of the spectrum and period spacing pattern of g-modes. Panel e: the sideways \'{e}chelle diagram of the g pattern. Panel f: the normalised sideways \'{e}chelle diagram of the g-modes pattern. Panels g to j: same as (c -- f) but for the r-mode patterns.  }\label{fig:KIC3240967}
\end{figure*}

We present the period spacing patterns of KIC\,3240967 as an example in Fig.~\ref{fig:KIC3240967}. 
KIC\,3240967 is the main-sequence $\gamma$\,Dor star with effective temperature of $7054\pm80\,\mathrm{K}$ \citep{Mathur_2017}. Its near-core rotation rate is $1.2857\pm0.0008\,\mathrm{d^{-1}}$ determined by the g- and r-modes pulsations (see section~\ref{subsec:TAR_fit_KIC3240967}).
 Panel (a) of Fig.~\ref{fig:KIC3240967} shows the periodogram from 0.2\,d to 2.4\,d, where the vertical dashed lines show the best fit from Eq.~\ref{equ:P_i} for each pattern. We found two period spacing patterns around 0.57\,d and 0.95\,d. The left pattern, marked by the blue dashed lines, comprises the dipole sectoral ($m=l=1$) g-modes while the right one, marked by the black dashed lines, comprises the even ($k=-2$) retrograde ($m=-1$) r-modes (for the geometry of r-modes, please refer to the Fig.2 in \cite{Saio_2018_Rossby_mode}). 
The mode identifications are made based on the TAR fit in Section~\ref{subsec:TAR_fit_KIC3240967} and previous literature \cite{VanReeth_2018} and \cite{Saio_2018_Rossby_mode}. 
Apart from these two patterns, the periodogram also displays two groups dominated by combination frequencies, located around 0.3\,d and 1.5\,d. The first is at twice the g-modes frequencies and the second is at the difference between the g and r-modes frequencies. 

Panel (b) in Fig.~\ref{fig:KIC3240967} depicts the period spacing versus period. The period spacing for g-modes decreases from 500\,s to 200\,s with increasing period. For the r-mode pattern, the period spacing increases from 200\,s to 1000\,s with increasing period. Both patterns show deviations from the linear model, such as the dip at 0.92\,d in the r-mode pattern, which is likely to be caused by the chemical composition gradient near the outer edge of the convective core \citep{Miglio_2008}. The period spacing in the r-mode pattern also shows a drop at the longest period spacings (see the right side of Fig.~\ref{fig:KIC3240967} (h) for example), caused by the rapid change of the eigenvalue $\lambda$ in the Laplace tidal equation (see Section~\ref{sec:near core rotation} for example).

Panels (c) and (d) zoom in on the g-modes from panels (a) and (b), while panels (g) and (h) do the same for the r-modes. In Panels (e) and (i), the \'{e}chelle diagrams are plotted sideways. The x-axis is the pulsation period while the y-axis is the term $P^{\mathrm{obs}}-n'\Delta P$ from the fit of eq.~\ref{equ:P_i}. For the peaks that do not belong to the pattern, we plotted them at the location that minimised the value $P^{\mathrm{obs}}-n'\Delta P$. Therefore, the y-axis reflects the deviations from the linear fit, similar to the curvature in the \'{e}chelle diagram of solar-like oscillators \citep[e.g.][]{Mazumdar_2014}. 
Panels (f) and (j) show the normalised sideways \'{e}chelle diagram. The x-axis is the index of peaks, counting the first peak as 0, and the y-axis is the deviation over the local period spacing $\left(P^{\mathrm{obs}}-n'\Delta P\right)/\Delta P$ expressed as a percentage.

\section{Near core rotation}\label{sec:near core rotation}

\subsection{TAR fitting algorithm}

We used the traditional approximation of rotation (TAR) to fit the period spacing patterns and derive the near-core rotation rate \citep{Eckart_1960, Lee1987, Townsend2005}. In the TAR, the $\theta$-component of the rotation vector and the effects of the centrifugal force are ignored \citep{Lee_1997_rotating}. The TAR saves computational time and it provides a reliable approximation of the influence of rotation on the pulsation periods. Many stars have been investigated using the TAR to determine the near-core rotation rates \citep[e.g.][]{VanReeth_2016_TAR, Guo2017, Saio_2018_Rossby_mode, Saio_2018} and we followed the method reported by \cite{VanReeth_2016_TAR}. We also assumed that all the stars rotate rigidly, consistent with recent observations \citep[e.g.][]{Kurtz_2014, Saio_2015, Murphy_2016, Schmid_2016, Guo2017, Aerts_2018, VanReeth_2018}. 

The spin parameter is defined as 
\begin{equation}
s\equiv\frac{2f_\mathrm{rot}}{f_\mathrm{co}},
\end{equation}
where $f_\mathrm{rot}$ is the rotation frequency and $f_\mathrm{co}$ is the pulsation frequency in the co-rotating frame. 
The spin parameter $s$ represents the validity of the TAR. \cite{Ballot2012} showed that the TAR remains valid for the pulsation frequencies of gravito-inertial modes with $s < 1$, even in the case of large stellar rotation rates. This was further studied by \cite{Ouazzani_2017}. The authors found that the TAR is well suited for the analysis of prograde and zonal modes when the stellar rotation is less than 50\,\% of the critical rotation rate, while retrograde gravito-inertial modes can be analysed with the TAR for rotation rates up to 25\,\% of the critical rotation rate. However, the TAR has also been used to determine the near-core rotation rate in moderate- to fast-rotating stars, by fitting observed r-modes and prograde gravito-inertial modes, with satisfactory results \cite[e.g.,][]{Saio_2018, VanReeth_2018}.

The pulsation period in the co-rotating frame, given by the TAR, is 
\begin{equation}
P^\mathrm{TAR}_{nlm, \mathrm{co}}=\frac{\Pi_0}{\sqrt{\lambda_{l,m,s}}}\left(n+ \varepsilon_g \right),  \label{equ:TAR_P}
\end{equation}
where $\Pi_0$ is the asymptotic period spacing, $n$ is the radial order, the phase term $\varepsilon_g$ is set as $0.5$, and $\lambda_{l,m,s}$ is the eigenvalue of the Laplace tidal equation, which is specified by the angular degree $l$ or the value $k$, the azimuthal order $m$ and the spin parameter $s$.

Eq.~\ref{equ:TAR_P} becomes the asymptotic relation \citep{Shibahashi1979} when $s\rightarrow 0$: 
\begin{equation}
P_{nl}=\frac{\Pi_0}{\sqrt{ l\left( l+1 \right)  }}\left(n+ \varepsilon_g \right). \label{equ:asymptotic spacing}
\end{equation}
The period in the inertial frame is obtained by 
\begin{equation}
P^\mathrm{TAR}_\mathrm{in}=\frac{1}{1/P^\mathrm{TAR}_\mathrm{co}+m\cdot f_\mathrm{rot}}. \label{equ:co_inertial}
\end{equation}
The period spacing is calculated with equ.~\ref{equ:TAR_P} and equ.~\ref{equ:co_inertial} to be
\begin{equation}
\Delta P^\mathrm{TAR}_{j, \mathrm{in}}=P^\mathrm{TAR}_{j+1, \mathrm{in}} - P^\mathrm{TAR}_{j, \mathrm{in}},
\end{equation}
where $j$ is the index sorted by the ascending period values. Note that the index $j$ is not the radial order since the radial order for r-modes drops with increasing period in the inertial reference frame.

The TAR fit can only give the slope, rather than the exact locations of peaks. Hence we interpolated the period spacing $\Delta P^\mathrm{interp}_i$ at the observational period $P_i^\mathrm{obs}$ using $\Delta P^\mathrm{TAR}$ and $P^\mathrm{TAR}$. We defined the log likelihood function $\ln L$ as 
\begin{equation}
\begin{split}
\ln L & \left(\Delta P | P^\mathrm{obs}, \Pi_0, f_\mathrm{rot}, k, m \right)\\
&=-\frac{1}{2} \sum_j \left[\frac{\Delta P^\mathrm{obs}_j-\Delta P^\mathrm{interp}_j}{\sigma_j^2}+\ln \left( 2\pi \sigma_j^2\right) \right], \label{equ:likelihood}
\end{split}
\end{equation}
where $\sigma$ is the residual of the best fitting result. Since the observed patterns showed glitches and dips caused by the chemical composition gradient that the TAR does not take into account, and these variations are larger than the formal error margins on the pulsation periods, we used the residuals rather than the period spacing uncertainties to estimate the TAR model uncertainty. The data points below $-1\sigma$ of the linear fit were excluded, because they were likely to be dips and mislead the fitting. We ran a Markov Chain Monto Carlo (MCMC) code to maximise the likelihood function Eq.~\ref{equ:likelihood} using the {\sc emcee} package \citep{Foreman-Mackey2013_emcee}. The eigenvalue $\lambda$ was given by {\sc gyre} \citep{Townsend_2013}. The fit has two variables because the value $k$ and the azimuthal order $m$ were given based on the mode identification, so only the asymptotic spacing $\Pi_0$ and the near-core rotation rate $f_\mathrm{rot}$ were free parameters. The initial value of $\Pi_0$ was 4200\,s, which is the most likely value of $\gamma$\,Dor stars from \cite{VanReeth_2016_TAR}. The initial guess of $f_\mathrm{rot}$ was chosen to be halfway between the prograde g and retrograde r patterns because the period of the prograde mode in the inertial frame is smaller than the rotation period, and vice versa for the r-modes. The range of $\Pi_0$ was from 2000 to 20000\,s to cover the typical range of both $\gamma$\,Dor and Slowly Pulsating B (SPB) stars. We searched for the best-fitting solution with a rotation rate $f_\mathrm{rot}$ between 0 and 5\,$\mathrm{d}^{-1}$. The upper limit of $f_\mathrm{rot}$ exceeds the critical rotation rate at which the centrifugal force causes a star to break apart. However, it is only selected to make sure that we can find a proper solution for every star. Based on our results (Fig.~\ref{fig:Pi0_rot} shown in Section~\ref{subsec:Asymptotic spacing and rotation}), there is no star which reaches this limit.
We also set the radial orders from 5 to 150. 

\subsection{TAR fitting of KIC\,3240967}\label{subsec:TAR_fit_KIC3240967}

\begin{figure}
\centering
\includegraphics[width=1\linewidth]{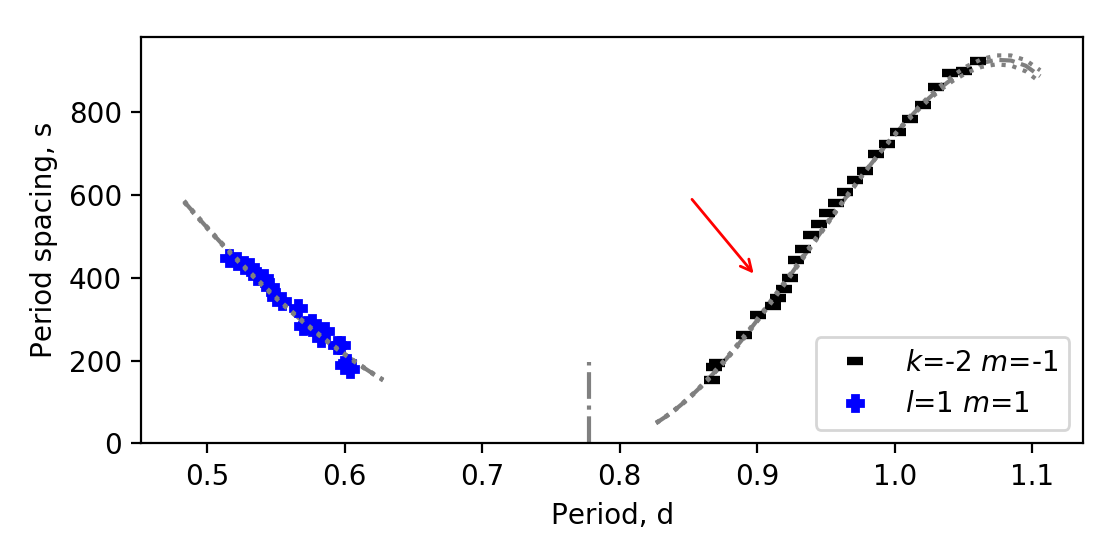}
\caption{The TAR fit of KIC\,3240967. The blue plus symbols show the $l=1, m=1$ gravity modes and the black minus symbols show the $k=-2, m=-1$ Rossby modes. The dashed grey curves display the best-fitted result. The error margins are plotted by the dotted lines, which are only visible at the right edge of the plot. The vertical dashed line denotes the fitted rotation period. The red arrow shows a dip. }\label{fig:KIC3240967_best_fit}
\end{figure}

\begin{figure}
\centering
\includegraphics[width=1\linewidth]{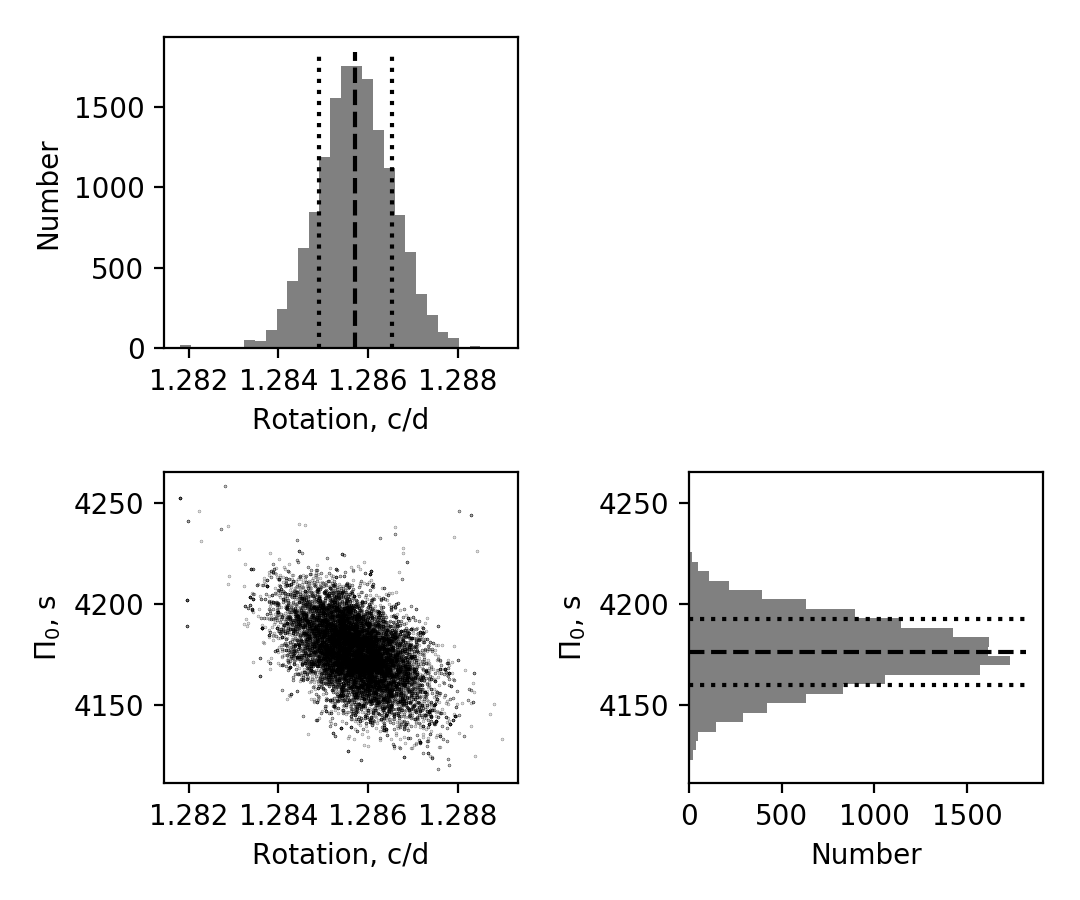}
\caption{The posterior distributions for the TAR fit to KIC\,3240967 using eq.~\ref{equ:likelihood}. The dashed lines are the medians and the dotted lines show the $\pm 1\sigma$ areas. }\label{fig:KIC3240967_posterior}
\end{figure}

\begin{figure}
\centering
\includegraphics[width=1\linewidth]{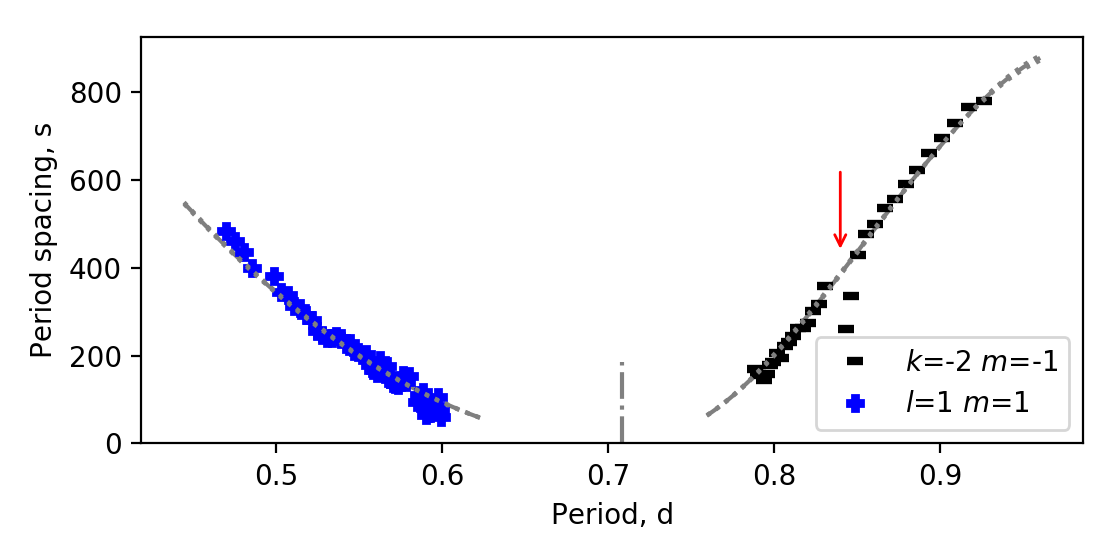}
\caption{Same as Fig.~\ref{fig:KIC3240967_best_fit} but for KIC\,11775251. An obvious dip are seen in r-modes pattern at 0.85\,d. Since the dip has been excluded, the fitting curves pass through other data points and do not be affected. The red arrow shows a dip. }\label{fig:KIC11775251_best_fit}
\end{figure}

We again use KIC\,3240967 to show how the TAR performed. Figure~\ref{fig:KIC3240967_best_fit} gives the TAR fit result. The dashed curves that pass through the data points represent the best model, while the dotted lines are the $\pm 1\sigma$ region. We found that the model can explain the observational patterns. The observed period spacings of r-modes follow the TAR prediction and the slope changes very little with period except for a drop near the end of the r pattern, which is caused by the rapid change of the eigenvalue $\lambda$ \citep[see the Figure 1 of both][]{Townsend_2003, Saio_2018_Rossby_mode}. The theoretical slope of the g-mode period spacing pattern also changes with the period but only slightly. A dip in r-modes appears at $P\sim 0.91\,\mathrm{d}$ (see the red arrow in Fig.~\ref{fig:KIC3240967_best_fit}), which might be the deviation from the asymptotic expression with the TAR, or is evidence of a chemical composition gradient effect \citep{Miglio_2008}. 

Figure~\ref{fig:KIC3240967_posterior} gives the posterior distributions of KIC\,3240967. The best-fitting rotation rate is $f_\mathrm{rot}=1.2857\pm0.0008\,\mathrm{d^{-1}}$ and the asymptotic spacing is $\Pi_0=4180\pm20\,\mathrm{s}$. Because the prograde and retrograde modes co-exist, the correlation between $f_\mathrm{rot}$ and $\Pi_0$ is weak, so we have a better constraint on the near-core rotation. 

Large dips are seen in some stars, for example, KIC\,11775251, and were omitted from the TAR fit. Figure~\ref{fig:KIC11775251_best_fit} shows the period spacing patterns and the TAR fit of KIC\,11775251, in which a dip of r-modes is seen around $\sim0.85\,\mathrm{d}$ denoted by the red arrow. Since the dip has been removed, the fitting curves pass through other data points and are not affected. The fitting results are: $\Pi_0=4040\pm20\,\mathrm{s}$ and $f_\mathrm{rot}=1.4115\pm0.0007\,\mathrm{d^{-1}}$.

\section{Period spacing patterns of \totalnumber $\gamma$\,Dor stars}\label{sec:observational_results}

\begin{table*} 
\caption{The mode identifications, mean pulsation periods, mean period spacings, slopes, asymptotic spacings, near-core rotation rates, and the minima and maxima of radial orders of \totalnumber stars in this paper.}\label{tab:rot_Pi0_table} 
\begin{tabular}{rrrrrrrrrr} 
\hline
KIC & $k$ & $m$ & Mean $P$ & Mean $\Delta P$ & mean $\Sigma$ & $\Pi_0$ & $f_\mathrm{rot}  $ & min $n$ & max $n$\\ 
    &     &     &   days   &    Seconds      &   days/days   & Seconds & $\mathrm{d^{-1}} $ &         &        \\ 
\hline
2575161 & \phantom{$-$}$0$ & \phantom{$-$}$1$ & 0.4 & 250 & \phantom{$-$}$-0.0466 \pm 0.0009$ & \phantom{$-$}$4470\pm 20$ & \phantom{$-$}$1.833\pm 0.001$\phantom{$0$} & 21 & 55\\
        & $-2$ & $-1$ & 0.7 & 580 & $0.0753 \pm 0.0004$ &         &         & 9 & 28\\
2710594 & \phantom{$-$}$0$ & \phantom{$-$}$1$ & 0.7 & 360 & \phantom{$-$}$-0.0290 \pm 0.0003$ & \phantom{$-$}$4000\pm 10$ & \phantom{$-$}$0.9920\pm 0.0006$ & 28 & 87\\
        & $-2$ & $-1$ & 1.2 & 470 & $0.0381 \pm 0.0005$ &         &         & 21 & 65\\
3240967 & \phantom{$-$}$0$ & \phantom{$-$}$1$ & 0.6 & 320 & \phantom{$-$}$-0.0344 \pm 0.0002$ & \phantom{$-$}$4180\pm 20$ & \phantom{$-$}$1.2857\pm 0.0008$ & 31 & 59\\
        & $-2$ & $-1$ & 0.9 & 450 & $0.0507 \pm 0.0004$ &         &         & 14 & 52\\
3341457 & \phantom{$-$}$0$ & \phantom{$-$}$1$ & 0.4 & 120 & \phantom{$-$}$-0.0292 \pm 0.0007$ & \phantom{$-$}$3840\pm 40$ & \phantom{$-$}$1.859\pm 0.001$\phantom{$0$} & 42 & 80\\
        & $-2$ & $-1$ & 0.6 & 100 & $0.0526 \pm 0.0008$ &         &         & 35 & 61\\
3445468 & \phantom{$-$}$0$ & \phantom{$-$}$1$ & 0.6 & 200 & \phantom{$-$}$-0.0266 \pm 0.0004$ & \phantom{$-$}$3860\pm 10$ & \phantom{$-$}$1.2037\pm 0.0007$ & 36 & 102\\
        & $-2$ & $-1$ & 1.0 & 560 & $0.0377 \pm 0.0006$ &         &         & 17 & 43\\
3448365 & \phantom{$-$}$0$ & \phantom{$-$}$1$ & 0.7 & 330 & \phantom{$-$}$-0.0301 \pm 0.0008$ & \phantom{$-$}$4190\pm 10$ & \phantom{$-$}$1.0736\pm 0.0005$ & 29 & 81\\
        & $-2$ & $-1$ & 1.1 & 380 & $0.0440 \pm 0.0003$ &         &         & 18 & 70\\
3449625 & \phantom{$-$}$0$ & \phantom{$-$}$2$ & 0.3 & 300 & \phantom{$-$}$-0.0317 \pm 0.0008$ & \phantom{$-$}$3970\pm 10$ & \phantom{$-$}$1.0514\pm 0.0005$ & 25 & 48\\
        & \phantom{$-$}$0$ & \phantom{$-$}$1$ & 0.7 & 380 & \phantom{$-$}$-0.0304 \pm 0.0002$ &         &         & 28 & 76\\
        & $-2$ & $-1$ & 1.1 & 310 & $0.049 \pm 0.001$\phantom{$0$} &         &         & 29 & 66\\
3626325 & \phantom{$-$}$0$ & \phantom{$-$}$1$ & 0.9 & 440 & \phantom{$-$}$-0.0248 \pm 0.0006$ & \phantom{$-$}$3910\pm 10$ & \phantom{$-$}$0.7820\pm 0.0005$ & 46 & 82\\
        & $-2$ & $-1$ & 1.6 & 690 & $0.0203 \pm 0.0006$ &         &         & 23 & 54\\
3868382 & \phantom{$-$}$0$ & \phantom{$-$}$2$ & 0.4 & 360 & \phantom{$-$}$-0.0325 \pm 0.0002$ & \phantom{$-$}$4450\pm 10$ & \phantom{$-$}$0.8441\pm 0.0004$ & 15 & 66\\
        & \phantom{$-$}$0$ & \phantom{$-$}$1$ & 0.7 & 630 & \phantom{$-$}$-0.0320 \pm 0.0003$ &         &         & 21 & 69\\
        & $-2$ & $-1$ & 1.5 & 530 & $0.0358 \pm 0.0002$ &         &         & 19 & 63\\
3942392 & \phantom{$-$}$0$ & \phantom{$-$}$2$ & 0.2 & 120 & \phantom{$-$}$-0.0402 \pm 0.0006$ & \phantom{$-$}$4270\pm 30$ & \phantom{$-$}$1.627\pm 0.001$\phantom{$0$} & 25 & 63\\
        & \phantom{$-$}$0$ & \phantom{$-$}$1$ & 0.5 & 230 & \phantom{$-$}$-0.0383 \pm 0.0003$ &         &         & 27 & 64\\
        & $-2$ & $-1$ & 0.7 & 120 & $0.0511 \pm 0.0005$ &         &         & 34 & 63\\
3971170 & \phantom{$-$}$0$ & \phantom{$-$}$1$ & 0.7 & 190 & \phantom{$-$}$-0.0264 \pm 0.0007$ & \phantom{$-$}$3920\pm 30$ & \phantom{$-$}$1.198\pm 0.001$\phantom{$0$} & 47 & 91\\
        & $-2$ & $-1$ & 1.0 & 240 & $0.066 \pm 0.001$\phantom{$0$} &         &         & 38 & 58\\
4069477 & \phantom{$-$}$0$ & \phantom{$-$}$1$ & 0.5 & 170 & \phantom{$-$}$-0.0328 \pm 0.0004$ & \phantom{$-$}$4020\pm 30$ & \phantom{$-$}$1.531\pm 0.002$\phantom{$0$} & 40 & 74\\
        & $-2$ & $-1$ & 0.8 & 640 & $0.0532 \pm 0.0006$ &         &         & 13 & 26\\
4076350 & \phantom{$-$}$0$ & \phantom{$-$}$2$ & 0.3 & 160 & \phantom{$-$}$-0.0583 \pm 0.0008$ & \phantom{$-$}$3470\pm 20$ & \phantom{$-$}$1.248\pm 0.001$\phantom{$0$} & 39 & 57\\
        & \phantom{$-$}$0$ & \phantom{$-$}$1$ & 0.6 & 230 & \phantom{$-$}$-0.0454 \pm 0.0008$ &         &         & 47 & 78\\
        & $-2$ & $-1$ & 1.0 & 430 & $0.0412 \pm 0.0007$ &         &         & 20 & 54\\
4077558 & \phantom{$-$}$0$ & \phantom{$-$}$1$ & 0.6 & 260 & \phantom{$-$}$-0.0340 \pm 0.0001$ & \phantom{$-$}$3960\pm 10$ & \phantom{$-$}$1.2063\pm 0.0008$ & 30 & 87\\
        & $-2$ & $-1$ & 1.1 & 620 & $0.0409 \pm 0.0006$ &         &         & 13 & 40\\
4261149 & \phantom{$-$}$0$ & \phantom{$-$}$1$ & 0.7 & 220 & \phantom{$-$}$-0.0209 \pm 0.0009$ & \phantom{$-$}$4820\pm 40$ & \phantom{$-$}$1.1251\pm 0.0009$ & 45 & 83\\
        & $-2$ & $-1$ & 1.1 & 440 & $0.024 \pm 0.004$\phantom{$0$} &         &         & 27 & 38\\
4448157 & \phantom{$-$}$0$ & \phantom{$-$}$1$ & 0.6 & 260 & \phantom{$-$}$-0.0303 \pm 0.0003$ & \phantom{$-$}$4000\pm 20$ & \phantom{$-$}$1.1698\pm 0.0008$ & 33 & 85\\
        & $-2$ & $-1$ & 1.0 & 200 & $0.0461 \pm 0.0007$ &         &         & 37 & 69\\
4566474 & \phantom{$-$}$0$ & \phantom{$-$}$1$ & 0.6 & 280 & \phantom{$-$}$-0.0312 \pm 0.0006$ & \phantom{$-$}$4130\pm 10$ & \phantom{$-$}$1.2345\pm 0.0007$ & 30 & 77\\
        & $-2$ & $-1$ & 1.0 & 570 & $0.0435 \pm 0.0004$ &         &         & 15 & 42\\
4585982 & \phantom{$-$}$0$ & \phantom{$-$}$1$ & 0.6 & 180 & \phantom{$-$}$-0.0249 \pm 0.0003$ & \phantom{$-$}$3370\pm 20$ & \phantom{$-$}$1.3924\pm 0.0009$ & 40 & 97\\
        & $-2$ & $-1$ & 0.8 & 130 & $0.030 \pm 0.001$\phantom{$0$} &         &         & 41 & 88\\
4672176 & \phantom{$-$}$0$ & \phantom{$-$}$1$ & 0.4 & 200 & \phantom{$-$}$-0.0398 \pm 0.0005$ & \phantom{$-$}$4280\pm 20$ & \phantom{$-$}$1.748\pm 0.001$\phantom{$0$} & 25 & 68\\
        & $-2$ & $-1$ & 0.7 & 290 & $0.0723 \pm 0.0004$ &         &         & 11 & 47\\
4758316 & \phantom{$-$}$0$ & \phantom{$-$}$1$ & 0.6 & 380 & \phantom{$-$}$-0.0395 \pm 0.0004$ & \phantom{$-$}$4470\pm 20$ & \phantom{$-$}$1.2445\pm 0.0009$ & 24 & 62\\
        & $-2$ & $-1$ & 1.0 & 820 & $0.042 \pm 0.001$\phantom{$0$} &         &         & 13 & 24\\
4774208 & \phantom{$-$}$0$ & \phantom{$-$}$1$ & 0.4 & 190 & \phantom{$-$}$-0.0427 \pm 0.0005$ & \phantom{$-$}$4340\pm 20$ & \phantom{$-$}$1.834\pm 0.001$\phantom{$0$} & 30 & 58\\
        & $-2$ & $-1$ & 0.6 & 290 & $0.0751 \pm 0.0004$ &         &         & 11 & 45\\
4843037 & \phantom{$-$}$0$ & \phantom{$-$}$1$ & 0.5 & 180 & \phantom{$-$}$-0.0320 \pm 0.0003$ & \phantom{$-$}$3910\pm 20$ & \phantom{$-$}$1.627\pm 0.001$\phantom{$0$} & 32 & 75\\
        & $-2$ & $-1$ & 0.7 & 330 & $0.0604 \pm 0.0004$ &         &         & 13 & 52\\
4857064 & \phantom{$-$}$0$ & \phantom{$-$}$1$ & 1.1 & 570 & \phantom{$-$}$-0.0206 \pm 0.0002$ & \phantom{$-$}$3832\pm 9$\phantom{$0$} & \phantom{$-$}$0.5738\pm 0.0004$ & 40 & 103\\
        & $-2$ & $-1$ & 2.3 & 740 & $0.010 \pm 0.001$\phantom{$0$} &         &         & 38 & 61\\
4859790 & \phantom{$-$}$0$ & \phantom{$-$}$1$ & 0.5 & 240 & \phantom{$-$}$-0.0358 \pm 0.0004$ & \phantom{$-$}$4550\pm 20$ & \phantom{$-$}$1.4584\pm 0.0007$ & 29 & 65\\
        & $-2$ & $-1$ & 0.8 & 320 & $0.0611 \pm 0.0004$ &         &         & 11 & 61\\
5040435 & \phantom{$-$}$0$ & \phantom{$-$}$1$ & 0.5 & 190 & \phantom{$-$}$-0.0296 \pm 0.0004$ & \phantom{$-$}$4070\pm 30$ & \phantom{$-$}$1.440\pm 0.001$\phantom{$0$} & 35 & 78\\
        & $-2$ & $-1$ & 0.8 & 280 & $0.073 \pm 0.004$\phantom{$0$} &         &         & 30 & 41\\
5114382 & \phantom{$-$}$0$ & \phantom{$-$}$1$ & 0.7 & 230 & \phantom{$-$}$-0.0234 \pm 0.0002$ & \phantom{$-$}$4110\pm 30$ & \phantom{$-$}$1.140\pm 0.001$\phantom{$0$} & 47 & 82\\
        & $-2$ & $-1$ & 1.1 & 510 & $0.057 \pm 0.002$\phantom{$0$} &         &         & 22 & 44\\
5294571 & \phantom{$-$}$0$ & \phantom{$-$}$2$ & 0.2 & 100 & \phantom{$-$}$-0.0318 \pm 0.0004$ & \phantom{$-$}$4150\pm 20$ & \phantom{$-$}$1.6421\pm 0.0009$ & 36 & 63\\
        & \phantom{$-$}$0$ & \phantom{$-$}$1$ & 0.5 & 200 & \phantom{$-$}$-0.0383 \pm 0.0005$ &         &         & 32 & 63\\
        & $-2$ & $-1$ & 0.7 & 250 & $0.0620 \pm 0.0004$ &         &         & 17 & 54\\
5391059 & \phantom{$-$}$0$ & \phantom{$-$}$1$ & 0.4 & 190 & \phantom{$-$}$-0.0419 \pm 0.0004$ & \phantom{$-$}$4210\pm 20$ & \phantom{$-$}$1.796\pm 0.001$\phantom{$0$} & 29 & 60\\
        & $-2$ & $-1$ & 0.7 & 340 & $0.0710 \pm 0.0002$ &         &         & 12 & 42\\
5476854 & \phantom{$-$}$0$ & \phantom{$-$}$1$ & 1.0 & 410 & \phantom{$-$}$-0.0183 \pm 0.0004$ & \phantom{$-$}$4060\pm 10$ & \phantom{$-$}$0.6647\pm 0.0004$ & 55 & 93\\
        & $-2$ & $-1$ & 1.9 & 630 & $0.0281 \pm 0.0006$ &         &         & 31 & 66\\
\hline
 \end{tabular} 
 \end{table*}

\setcounter{table}{0}

\begin{table*} 
\caption{continued.}\label{tab:rot_Pi0_table_2} 
\begin{tabular}{rrrrrrrrrr} 
\hline
KIC & $k$ & $m$ & Mean $P$ & Mean $\Delta P$ & mean $\Sigma$ & $\Pi_0$ & $f_\mathrm{rot}  $ & min $n$ & max $n$\\ 
    &     &     &   days   &    Seconds      &   days/days   & Seconds & $\mathrm{d^{-1}} $ &         &        \\ 
\hline
5640438 & \phantom{$-$}$0$ & \phantom{$-$}$1$ & 0.7 & 370 & \phantom{$-$}$-0.0276 \pm 0.0003$ & \phantom{$-$}$4230\pm 20$ & \phantom{$-$}$0.9788\pm 0.0005$ & 36 & 74\\
        & $-2$ & $-1$ & 1.2 & 390 & $0.0376 \pm 0.0003$ &         &         & 15 & 82\\
5706866 & \phantom{$-$}$0$ & \phantom{$-$}$1$ & 0.6 & 180 & \phantom{$-$}$-0.027 \pm 0.001$\phantom{$0$} & \phantom{$-$}$3610\pm 20$ & \phantom{$-$}$1.218\pm 0.001$\phantom{$0$} & 61 & 91\\
        & $-2$ & $-1$ & 1.1 & 610 & $0.0342 \pm 0.0004$ &         &         & 16 & 38\\
5708550 & \phantom{$-$}$0$ & \phantom{$-$}$1$ & 0.9 & 360 & \phantom{$-$}$-0.0233 \pm 0.0004$ & \phantom{$-$}$3840\pm 10$ & \phantom{$-$}$0.7911\pm 0.0006$ & 41 & 100\\
        & $-2$ & $-1$ & 1.7 & 800 & $0.010 \pm 0.001$\phantom{$0$} &         &         & 17 & 41\\
5721632 & \phantom{$-$}$0$ & \phantom{$-$}$1$ & 0.4 & 320 & \phantom{$-$}$-0.0479 \pm 0.0005$ & \phantom{$-$}$4371\pm 8$\phantom{$0$} & \phantom{$-$}$1.6969\pm 0.0009$ & 18 & 54\\
        & $-2$ & $-1$ & 0.7 & 220 & $0.0696 \pm 0.0004$ &         &         & 25 & 44\\
        & $-1$ & $-1$ & 1.9 & 4270 & $0.094 \pm 0.001$\phantom{$0$} &         &         & 25 & 43\\
5801556 & \phantom{$-$}$0$ & \phantom{$-$}$2$ & 0.3 & 230 & \phantom{$-$}$-0.0401 \pm 0.0008$ & \phantom{$-$}$4240\pm 20$ & \phantom{$-$}$1.2186\pm 0.0008$ & 20 & 56\\
        & \phantom{$-$}$0$ & \phantom{$-$}$1$ & 0.6 & 230 & \phantom{$-$}$-0.042 \pm 0.002$\phantom{$0$} &         &         & 44 & 70\\
        & $-2$ & $-1$ & 1.1 & 740 & $0.0335 \pm 0.0006$ &         &         & 13 & 30\\
5984615 & \phantom{$-$}$0$ & \phantom{$-$}$1$ & 0.8 & 240 & \phantom{$-$}$-0.0235 \pm 0.0002$ & \phantom{$-$}$3820\pm 20$ & \phantom{$-$}$0.9892\pm 0.0007$ & 48 & 102\\
        & $-2$ & $-1$ & 1.3 & 530 & $0.0368 \pm 0.0008$ &         &         & 21 & 62\\
6048255 & \phantom{$-$}$0$ & \phantom{$-$}$1$ & 0.6 & 170 & \phantom{$-$}$-0.0252 \pm 0.0002$ & \phantom{$-$}$3750\pm 30$ & \phantom{$-$}$1.3748\pm 0.0009$ & 44 & 95\\
        & $-2$ & $-1$ & 0.8 & 130 & $0.0374 \pm 0.0005$ &         &         & 41 & 77\\
6291473 & \phantom{$-$}$0$ & \phantom{$-$}$2$ & 0.2 & 110 & \phantom{$-$}$-0.0329 \pm 0.0004$ & \phantom{$-$}$5770\pm 40$ & \phantom{$-$}$1.671\pm 0.001$\phantom{$0$} & 21 & 64\\
        & \phantom{$-$}$0$ & \phantom{$-$}$1$ & 0.5 & 210 & \phantom{$-$}$-0.0342 \pm 0.0004$ &         &         & 27 & 58\\
        & $-2$ & $-1$ & 0.7 & 310 & $0.11 \pm 0.02$\phantom{$0$}\phantom{$0$} &         &         & 23 & 30\\
6301745 & \phantom{$-$}$0$ & \phantom{$-$}$1$ & 0.3 & 210 & \phantom{$-$}$-0.0495 \pm 0.0007$ & \phantom{$-$}$4210\pm 40$ & \phantom{$-$}$2.297\pm 0.002$\phantom{$0$} & 23 & 43\\
        & $-2$ & $-1$ & 0.5 & 110 & $0.0672 \pm 0.0008$ &         &         & 23 & 48\\
6468987 & \phantom{$-$}$0$ & \phantom{$-$}$1$ & 0.5 & 230 & \phantom{$-$}$-0.035 \pm 0.001$\phantom{$0$} & \phantom{$-$}$4020\pm 50$ & \phantom{$-$}$1.551\pm 0.002$\phantom{$0$} & 30 & 65\\
        & $-2$ & $-1$ & 0.7 & 190 & $0.07 \pm 0.01$\phantom{$0$}\phantom{$0$} &         &         & 38 & 46\\
6696689 & \phantom{$-$}$0$ & \phantom{$-$}$1$ & 0.7 & 180 & \phantom{$-$}$-0.0239 \pm 0.0002$ & \phantom{$-$}$3830\pm 20$ & \phantom{$-$}$1.1057\pm 0.0006$ & 52 & 105\\
        & $-2$ & $-1$ & 1.0 & 190 & $0.0347 \pm 0.0003$ &         &         & 45 & 80\\
6806005 & \phantom{$-$}$0$ & \phantom{$-$}$1$ & 0.6 & 240 & \phantom{$-$}$-0.0307 \pm 0.0006$ & \phantom{$-$}$4000\pm 20$ & \phantom{$-$}$1.3235\pm 0.0008$ & 35 & 72\\
        & $-2$ & $-1$ & 0.9 & 380 & $0.0501 \pm 0.0004$ &         &         & 14 & 61\\
6923424 & \phantom{$-$}$0$ & \phantom{$-$}$2$ & 0.2 & 80 & \phantom{$-$}$-0.0398 \pm 0.0005$ & \phantom{$-$}$4230\pm 30$ & \phantom{$-$}$2.023\pm 0.002$\phantom{$0$} & 23 & 75\\
        & \phantom{$-$}$0$ & \phantom{$-$}$1$ & 0.4 & 240 & \phantom{$-$}$-0.0496 \pm 0.0006$ &         &         & 17 & 56\\
        & $-2$ & $-1$ & 0.5 & 80 & $0.0494 \pm 0.0006$ &         &         & 32 & 60\\
7039007 & \phantom{$-$}$0$ & \phantom{$-$}$1$ & 0.7 & 290 & \phantom{$-$}$-0.0300 \pm 0.0005$ & \phantom{$-$}$3970\pm 20$ & \phantom{$-$}$1.039\pm 0.001$\phantom{$0$} & 46 & 75\\
        & $-2$ & $-1$ & 1.2 & 650 & $0.0257 \pm 0.0007$ &         &         & 21 & 36\\
7137351 & \phantom{$-$}$0$ & \phantom{$-$}$1$ & 0.6 & 230 & \phantom{$-$}$-0.0329 \pm 0.0003$ & \phantom{$-$}$3820\pm 10$ & \phantom{$-$}$1.2129\pm 0.0009$ & 36 & 89\\
        & $-2$ & $-1$ & 1.0 & 550 & $0.0353 \pm 0.0003$ &         &         & 16 & 48\\
7436266 & \phantom{$-$}$0$ & \phantom{$-$}$1$ & 0.6 & 220 & \phantom{$-$}$-0.0298 \pm 0.0003$ & \phantom{$-$}$4070\pm 10$ & \phantom{$-$}$1.2897\pm 0.0007$ & 32 & 91\\
        & $-2$ & $-1$ & 0.9 & 370 & $0.0487 \pm 0.0006$ &         &         & 10 & 68\\
7538181 & \phantom{$-$}$0$ & \phantom{$-$}$1$ & 0.6 & 260 & \phantom{$-$}$-0.0335 \pm 0.0008$ & \phantom{$-$}$3840\pm 20$ & \phantom{$-$}$1.1939\pm 0.0009$ & 41 & 73\\
        & $-2$ & $-1$ & 1.1 & 580 & $0.0405 \pm 0.0009$ &         &         & 14 & 41\\
7551589 & \phantom{$-$}$0$ & \phantom{$-$}$1$ & 0.6 & 180 & \phantom{$-$}$-0.0292 \pm 0.0003$ & \phantom{$-$}$3820\pm 20$ & \phantom{$-$}$1.3182\pm 0.0009$ & 43 & 91\\
        & $-2$ & $-1$ & 0.9 & 490 & $0.0460 \pm 0.0008$ &         &         & 15 & 46\\
7583663 & \phantom{$-$}$0$ & \phantom{$-$}$1$ & 0.6 & 280 & \phantom{$-$}$-0.0307 \pm 0.0005$ & \phantom{$-$}$4220\pm 20$ & \phantom{$-$}$1.1608\pm 0.0007$ & 38 & 71\\
        & $-2$ & $-1$ & 1.0 & 180 & $0.0385 \pm 0.0005$ &         &         & 39 & 70\\
7596250 & \phantom{$-$}$0$ & \phantom{$-$}$1$ & 0.6 & 350 & \phantom{$-$}$-0.0346 \pm 0.0005$ & \phantom{$-$}$4290\pm 20$ & \phantom{$-$}$1.1876\pm 0.0007$ & 28 & 67\\
        & $-2$ & $-1$ & 1.0 & 590 & $0.0502 \pm 0.0006$ &         &         & 17 & 38\\
7621649 & \phantom{$-$}$0$ & \phantom{$-$}$1$ & 0.8 & 510 & \phantom{$-$}$-0.0270 \pm 0.0002$ & \phantom{$-$}$3990\pm 10$ & \phantom{$-$}$0.7745\pm 0.0004$ & 31 & 81\\
        & $-2$ & $-1$ & 1.6 & 530 & $0.0307 \pm 0.0004$ &         &         & 29 & 63\\
7778826 & \phantom{$-$}$0$ & \phantom{$-$}$1$ & 0.4 & 270 & \phantom{$-$}$-0.0510 \pm 0.0004$ & \phantom{$-$}$4020\pm 30$ & \phantom{$-$}$2.060\pm 0.002$\phantom{$0$} & 21 & 48\\
        & $-2$ & $-1$ & 0.6 & 330 & $0.0792 \pm 0.0007$ &         &         & 9 & 41\\
8198031 & \phantom{$-$}$0$ & \phantom{$-$}$1$ & 0.4 & 170 & \phantom{$-$}$-0.0375 \pm 0.0005$ & \phantom{$-$}$4140\pm 20$ & \phantom{$-$}$1.855\pm 0.001$\phantom{$0$} & 26 & 75\\
        & $-2$ & $-1$ & 0.6 & 370 & $0.0767 \pm 0.0006$ &         &         & 13 & 39\\
8263970 & \phantom{$-$}$0$ & \phantom{$-$}$1$ & 0.4 & 170 & \phantom{$-$}$-0.0393 \pm 0.0007$ & \phantom{$-$}$4180\pm 20$ & \phantom{$-$}$1.946\pm 0.001$\phantom{$0$} & 22 & 77\\
        & $-2$ & $-1$ & 0.6 & 470 & $0.0786 \pm 0.0001$ &         &         & 11 & 30\\
8316105 & \phantom{$-$}$0$ & \phantom{$-$}$1$ & 0.6 & 300 & \phantom{$-$}$-0.0354 \pm 0.0005$ & \phantom{$-$}$4240\pm 20$ & \phantom{$-$}$1.3352\pm 0.0008$ & 25 & 71\\
        & $-2$ & $-1$ & 0.9 & 420 & $0.0539 \pm 0.0004$ &         &         & 16 & 51\\
8330056 & \phantom{$-$}$0$ & \phantom{$-$}$1$ & 0.4 & 170 & \phantom{$-$}$-0.0393 \pm 0.0005$ & \phantom{$-$}$4100\pm 20$ & \phantom{$-$}$1.913\pm 0.001$\phantom{$0$} & 27 & 69\\
        & $-2$ & $-1$ & 0.6 & 390 & $0.0731 \pm 0.0006$ &         &         & 10 & 43\\
8351778 & \phantom{$-$}$0$ & \phantom{$-$}$2$ & 0.3 & 260 & \phantom{$-$}$-0.0361 \pm 0.0007$ & \phantom{$-$}$3760\pm 20$ & \phantom{$-$}$0.9645\pm 0.0009$ & 30 & 58\\
        & \phantom{$-$}$0$ & \phantom{$-$}$1$ & 0.7 & 320 & \phantom{$-$}$-0.0262 \pm 0.0002$ &         &         & 41 & 85\\
        & $-2$ & $-1$ & 1.2 & 130 & $0.026 \pm 0.002$\phantom{$0$} &         &         & 72 & 86\\
8375138 & \phantom{$-$}$0$ & \phantom{$-$}$1$ & 0.5 & 200 & \phantom{$-$}$-0.0376 \pm 0.0003$ & \phantom{$-$}$4150\pm 20$ & \phantom{$-$}$1.6421\pm 0.0009$ & 30 & 66\\
        & $-2$ & $-1$ & 0.7 & 400 & $0.0659 \pm 0.0003$ &         &         & 9 & 43\\
8973529 & \phantom{$-$}$0$ & \phantom{$-$}$2$ & 0.3 & 140 & \phantom{$-$}$-0.0308 \pm 0.0003$ & \phantom{$-$}$4000\pm 20$ & \phantom{$-$}$1.204\pm 0.001$\phantom{$0$} & 23 & 91\\
        & \phantom{$-$}$0$ & \phantom{$-$}$1$ & 0.6 & 320 & \phantom{$-$}$-0.0325 \pm 0.0004$ &         &         & 23 & 84\\
        & $-2$ & $-1$ & 0.9 & 220 & $0.066 \pm 0.002$\phantom{$0$} &         &         & 39 & 60\\
\hline
 \end{tabular} 
 \end{table*}

\setcounter{table}{0}

\begin{table*} 
\caption{continued.}\label{tab:rot_Pi0_table_3} 
\begin{tabular}{rrrrrrrrrr} 
\hline
KIC & $k$ & $m$ & Mean $P$ & Mean $\Delta P$ & mean $\Sigma$ & $\Pi_0$ & $f_\mathrm{rot}  $ & min $n$ & max $n$\\ 
    &     &     &   days   &    Seconds      &   days/days   & Seconds & $\mathrm{d^{-1}} $ &         &        \\ 
\hline
9210943 & \phantom{$-$}$0$ & \phantom{$-$}$1$ & 0.4 & 300 & \phantom{$-$}$-0.043 \pm 0.001$\phantom{$0$} & \phantom{$-$}$4310\pm 20$ & \phantom{$-$}$1.703\pm 0.001$\phantom{$0$} & 25 & 46\\
        & $-2$ & $-1$ & 0.7 & 410 & $0.0693 \pm 0.0005$ &         &         & 10 & 41\\
9419182 & \phantom{$-$}$0$ & \phantom{$-$}$2$ & 0.3 & 180 & \phantom{$-$}$-0.0385 \pm 0.0006$ & \phantom{$-$}$4190\pm 10$ & \phantom{$-$}$1.1510\pm 0.0006$ & 28 & 63\\
        & \phantom{$-$}$0$ & \phantom{$-$}$1$ & 0.7 & 250 & \phantom{$-$}$-0.0282 \pm 0.0003$ &         &         & 38 & 83\\
        & $-2$ & $-1$ & 1.1 & 470 & $0.0434 \pm 0.0006$ &         &         & 16 & 62\\
9480469 & \phantom{$-$}$0$ & \phantom{$-$}$1$ & 0.5 & 160 & \phantom{$-$}$-0.032 \pm 0.001$\phantom{$0$} & \phantom{$-$}$4330\pm 20$ & \phantom{$-$}$1.554\pm 0.001$\phantom{$0$} & 41 & 73\\
        & $-2$ & $-1$ & 0.8 & 400 & $0.0647 \pm 0.0002$ &         &         & 12 & 45\\
9594007 & \phantom{$-$}$0$ & \phantom{$-$}$2$ & 0.2 & 130 & \phantom{$-$}$-0.0428 \pm 0.0005$ & \phantom{$-$}$3980\pm 20$ & \phantom{$-$}$1.671\pm 0.001$\phantom{$0$} & 28 & 54\\
        & \phantom{$-$}$0$ & \phantom{$-$}$1$ & 0.5 & 200 & \phantom{$-$}$-0.0366 \pm 0.0008$ &         &         & 23 & 81\\
        & $-2$ & $-1$ & 0.7 & 160 & $0.0550 \pm 0.0007$ &         &         & 26 & 62\\
9594100 & \phantom{$-$}$0$ & \phantom{$-$}$2$ & 0.3 & 240 & \phantom{$-$}$-0.0370 \pm 0.0005$ & \phantom{$-$}$4110\pm 20$ & \phantom{$-$}$1.2159\pm 0.0009$ & 24 & 48\\
        & \phantom{$-$}$0$ & \phantom{$-$}$1$ & 0.6 & 260 & \phantom{$-$}$-0.0313 \pm 0.0003$ &         &         & 32 & 78\\
        & $-2$ & $-1$ & 1.0 & 320 & $0.03 \pm 0.01$\phantom{$0$}\phantom{$0$} &         &         & 32 & 47\\
9652302 & \phantom{$-$}$0$ & \phantom{$-$}$1$ & 0.8 & 330 & \phantom{$-$}$-0.021 \pm 0.001$\phantom{$0$} & \phantom{$-$}$3710\pm 20$ & \phantom{$-$}$0.9147\pm 0.0006$ & 49 & 89\\
        & $-2$ & $-1$ & 1.3 & 270 & $0.061 \pm 0.003$\phantom{$0$} &         &         & 48 & 67\\
9716563 & \phantom{$-$}$0$ & \phantom{$-$}$2$ & 0.4 & 210 & \phantom{$-$}$-0.0258 \pm 0.0002$ & \phantom{$-$}$3860\pm 10$ & \phantom{$-$}$0.9081\pm 0.0008$ & 30 & 84\\
        & \phantom{$-$}$0$ & \phantom{$-$}$1$ & 0.7 & 570 & \phantom{$-$}$-0.0307 \pm 0.0003$ &         &         & 24 & 68\\
        & $-2$ & $-1$ & 1.5 & 820 & $0.0119 \pm 0.0007$ &         &         & 18 & 27\\
9962653 & \phantom{$-$}$0$ & \phantom{$-$}$1$ & 0.4 & 200 & \phantom{$-$}$-0.0410 \pm 0.0004$ & \phantom{$-$}$4150\pm 20$ & \phantom{$-$}$1.763\pm 0.001$\phantom{$0$} & 28 & 62\\
        & $-2$ & $-1$ & 0.7 & 290 & $0.0686 \pm 0.0002$ &         &         & 11 & 50\\
10423501 & \phantom{$-$}$0$ & \phantom{$-$}$1$ & 0.8 & 480 & \phantom{$-$}$-0.0271 \pm 0.0002$ & \phantom{$-$}$4240\pm 20$ & \phantom{$-$}$0.8420\pm 0.0006$ & 30 & 77\\
        & $-2$ & $-1$ & 1.4 & 480 & $0.0339 \pm 0.0009$ &         &         & 37 & 51\\
10481462 & \phantom{$-$}$0$ & \phantom{$-$}$1$ & 0.4 & 210 & \phantom{$-$}$-0.052 \pm 0.001$\phantom{$0$} & \phantom{$-$}$4410\pm 30$ & \phantom{$-$}$2.208\pm 0.002$\phantom{$0$} & 18 & 53\\
        & $-2$ & $-1$ & 0.5 & 440 & $0.0973 \pm 0.0005$ &         &         & 11 & 26\\
10818266 & \phantom{$-$}$0$ & \phantom{$-$}$1$ & 0.4 & 250 & \phantom{$-$}$-0.0468 \pm 0.0005$ & \phantom{$-$}$4290\pm 20$ & \phantom{$-$}$1.849\pm 0.001$\phantom{$0$} & 19 & 57\\
        & $-2$ & $-1$ & 0.7 & 460 & $0.0772 \pm 0.0003$ &         &         & 11 & 34\\
10859386 & \phantom{$-$}$0$ & \phantom{$-$}$1$ & 0.4 & 290 & \phantom{$-$}$-0.045 \pm 0.001$\phantom{$0$} & \phantom{$-$}$4420\pm 30$ & \phantom{$-$}$1.813\pm 0.001$\phantom{$0$} & 24 & 45\\
        & $-2$ & $-1$ & 0.6 & 120 & $0.062 \pm 0.002$\phantom{$0$} &         &         & 35 & 52\\
11256244 & \phantom{$-$}$0$ & \phantom{$-$}$1$ & 0.8 & 300 & \phantom{$-$}$-0.0242 \pm 0.0002$ & \phantom{$-$}$3930\pm 10$ & \phantom{$-$}$0.9622\pm 0.0006$ & 46 & 85\\
        & $-2$ & $-1$ & 1.3 & 640 & $0.026 \pm 0.001$\phantom{$0$} &         &         & 23 & 45\\
11466960 & \phantom{$-$}$0$ & \phantom{$-$}$1$ & 0.6 & 210 & \phantom{$-$}$-0.0324 \pm 0.0005$ & \phantom{$-$}$4240\pm 10$ & \phantom{$-$}$1.3874\pm 0.0007$ & 30 & 83\\
        & $-2$ & $-1$ & 0.9 & 400 & $0.0560 \pm 0.0004$ &         &         & 13 & 53\\
11519475 & \phantom{$-$}$0$ & \phantom{$-$}$1$ & 0.5 & 190 & \phantom{$-$}$-0.0333 \pm 0.0003$ & \phantom{$-$}$4070\pm 30$ & \phantom{$-$}$1.576\pm 0.001$\phantom{$0$} & 30 & 76\\
        & $-2$ & $-1$ & 0.7 & 200 & $0.0550 \pm 0.0008$ &         &         & 31 & 50\\
11550154 & \phantom{$-$}$0$ & \phantom{$-$}$2$ & 0.2 & 90 & \phantom{$-$}$-0.0409 \pm 0.0003$ & \phantom{$-$}$4140\pm 20$ & \phantom{$-$}$2.017\pm 0.001$\phantom{$0$} & 25 & 59\\
        & \phantom{$-$}$0$ & \phantom{$-$}$1$ & 0.4 & 140 & \phantom{$-$}$-0.0367 \pm 0.0006$ &         &         & 24 & 80\\
        & $-2$ & $-1$ & 0.6 & 320 & $0.0794 \pm 0.0005$ &         &         & 11 & 40\\
11571757 & \phantom{$-$}$0$ & \phantom{$-$}$2$ & 0.4 & 220 & \phantom{$-$}$-0.048 \pm 0.002$\phantom{$0$} & \phantom{$-$}$6130\pm 30$ & \phantom{$-$}$1.0291\pm 0.0005$ & 29 & 52\\
        & \phantom{$-$}$0$ & \phantom{$-$}$1$ & 0.8 & 240 & \phantom{$-$}$-0.0223 \pm 0.0004$ &         &         & 32 & 105\\
        & $-2$ & $-1$ & 1.1 & 380 & $0.064 \pm 0.002$\phantom{$0$} &         &         & 26 & 46\\
11657371 & \phantom{$-$}$0$ & \phantom{$-$}$1$ & 0.4 & 250 & \phantom{$-$}$-0.0436 \pm 0.0007$ & \phantom{$-$}$3900\pm 40$ & \phantom{$-$}$1.848\pm 0.002$\phantom{$0$} & 27 & 52\\
        & $-2$ & $-1$ & 0.6 & 60 & $0.077 \pm 0.002$\phantom{$0$} &         &         & 50 & 71\\
11775251 & \phantom{$-$}$0$ & \phantom{$-$}$1$ & 0.6 & 190 & \phantom{$-$}$-0.0327 \pm 0.0003$ & \phantom{$-$}$4040\pm 20$ & \phantom{$-$}$1.4115\pm 0.0007$ & 32 & 89\\
        & $-2$ & $-1$ & 0.8 & 350 & $0.0572 \pm 0.0006$ &         &         & 15 & 49\\
11907454 & \phantom{$-$}$0$ & \phantom{$-$}$1$ & 0.6 & 240 & \phantom{$-$}$-0.0368 \pm 0.0006$ & \phantom{$-$}$4200\pm 20$ & \phantom{$-$}$1.3387\pm 0.0006$ & 31 & 75\\
        & $-2$ & $-1$ & 0.9 & 270 & $0.0543 \pm 0.0001$ &         &         & 11 & 67\\
12066947 & \phantom{$-$}$0$ & \phantom{$-$}$1$ & 0.4 & 140 & \phantom{$-$}$-0.0399 \pm 0.0009$ & \phantom{$-$}$4170\pm 30$ & \phantom{$-$}$2.159\pm 0.002$\phantom{$0$} & 26 & 65\\
        & $-2$ & $-1$ & 0.5 & 270 & $0.0804 \pm 0.0005$ &         &         & 12 & 45\\
12170722 & \phantom{$-$}$0$ & \phantom{$-$}$2$ & 0.2 & 100 & \phantom{$-$}$-0.0362 \pm 0.0004$ & \phantom{$-$}$4270\pm 20$ & \phantom{$-$}$1.702\pm 0.001$\phantom{$0$} & 23 & 78\\
        & \phantom{$-$}$0$ & \phantom{$-$}$1$ & 0.5 & 150 & \phantom{$-$}$-0.0334 \pm 0.0002$ &         &         & 34 & 80\\
        & $-2$ & $-1$ & 0.7 & 390 & $0.0715 \pm 0.0005$ &         &         & 11 & 39\\
12303838 & \phantom{$-$}$0$ & \phantom{$-$}$1$ & 0.6 & 250 & \phantom{$-$}$-0.0329 \pm 0.0004$ & \phantom{$-$}$4160\pm 20$ & \phantom{$-$}$1.3301\pm 0.0007$ & 34 & 71\\
        & $-2$ & $-1$ & 0.8 & 190 & $0.0483 \pm 0.0005$ &         &         & 28 & 69\\
12520187 & \phantom{$-$}$0$ & \phantom{$-$}$1$ & 0.4 & 310 & \phantom{$-$}$-0.0568 \pm 0.0006$ & \phantom{$-$}$4650\pm 50$ & \phantom{$-$}$1.872\pm 0.002$\phantom{$0$} & 19 & 45\\
        & $-2$ & $-1$ & 0.6 & 90 & $0.062 \pm 0.003$\phantom{$0$} &         &         & 30 & 52\\
\hline
 \end{tabular} 
 \end{table*}

We inspected the Fourier transforms of stellar light curves for 1593 \textit{Kepler} targets with effective temperatures from 6600\,K to 10000\,K. Among these, we found \totalnumber stars which show resolved period spacing patterns of both g-modes and r-modes. Table~\ref{tab:rot_Pi0_table} shows mode identifications, observational pattern parameters, asymptotic spacings, near-core rotation rates, and the radial order ranges for our sample. The means of the periods are the average of all detected peaks. The mean period spacing is the slope of the linear fit between $P_i$ and the index $i$, which is similar to the method used by \cite{White_2011} to find p-mode frequency spacings $\left(\Delta \nu \right)$ for solar-like oscillators. 
In Appendix~\ref{appendix: period spacing diagrams}, we show the equivalent of Figs.~\ref{fig:KIC3240967} to \ref{fig:KIC3240967_posterior} for all \totalnumber stars with g and r-modes, sorted by rotation rate. The axes of panel (a) and (b) are the same for all figures, to make the comparison straightforward. Example diagrams are shown in figs~\ref{appfig:KIC6301745} to \ref{appfig:KIC6301745_posterior}. The figures for the other \totalnumberminusone stars are available as supplementary online material. 
Every star shows both downward and upward period spacing patterns, which are $l=1,m=1$ g-modes and $k=-2, m=-1$ r-modes, respectively. As the rotation decreases, both the g- and r-mode patterns move from left to right and the period gap between them increases. We also found \quadrupolemodenumber patterns which are identified as the $l=2, m=2$ g-modes (listed in Table~\ref{tab:rot_Pi0_table}). Only one star (KIC\,5721632) has $k=-1, m=-1$ r-modes. 

\subsection{Stars on the HR diagram}

\begin{figure}
\centering
\includegraphics[width=\linewidth]{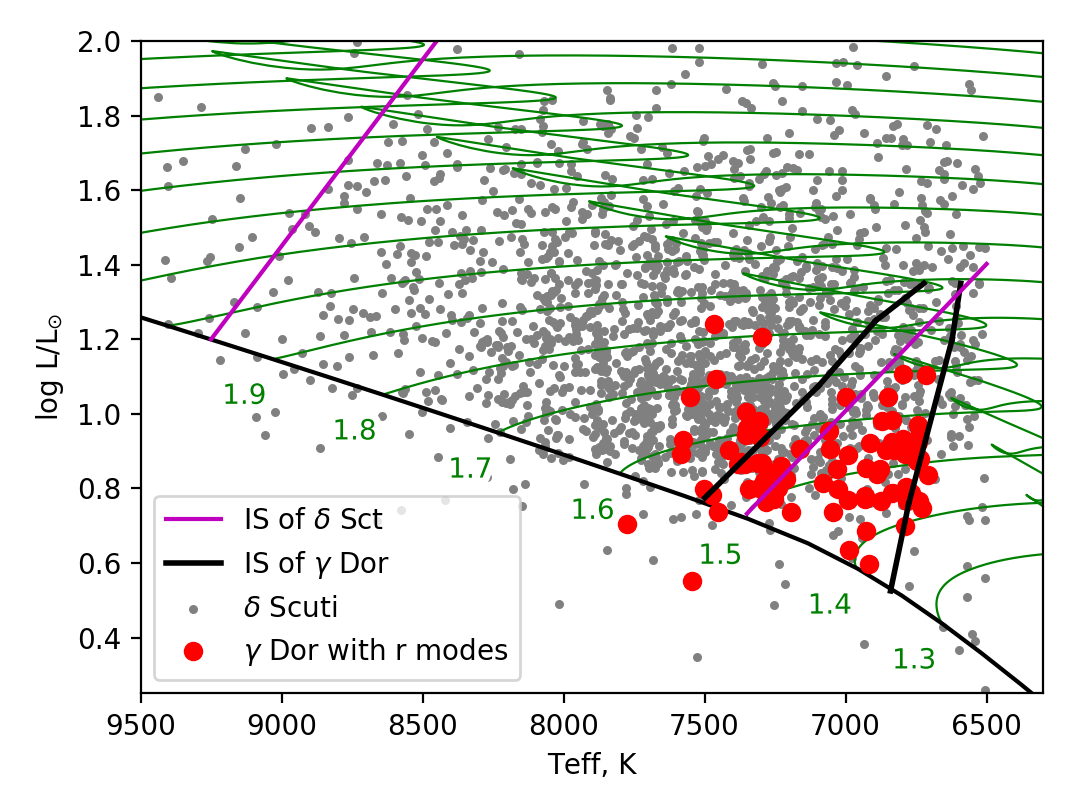}
\caption{The $\delta$\,Scuti and $\gamma$\,Doradus stars on the HR diagram. The grey dashed dots are the $\delta$\,Scuti stars. The red circles are the \totalnumberminusfour $\gamma$\,Dor stars with Rossby modes in this paper. The green numbers are the masses of the evolutionary tracks. The instability strips of $\delta$\,Scuti and $\gamma$\,Dor are shown by the purple and black solid lines. }\label{fig:HR_diagram}
\end{figure}

Figure~\ref{fig:HR_diagram} shows the HR diagram of the \totalnumberminusfour $\gamma$\,Dor stars (red circles) with clear r-mode patterns. \excludednumberHRdiagram stars are excluded since their data quality are insufficient. Our exclusion conditions are: (1) the fractional parallax uncertainty is larger than 20\%; (2) the star is 0.4\,dex below the zero-age main-sequence (ZAMS); (3) there is no parallax or parallax error in \textit{Gaia} DR2; (4) the star is identified as a binary. 
We took the effective temperatures $T_\mathrm{eff}$ from the latest version of the \textit{Kepler} Stellar Properties Catalog \citep[KSPC DR25,][]{Mathur_2017}. The luminosity $L$ was calculated by \cite{Murphy_2019} using the \textit{Gaia} Data Release 2 (DR2) parallax \citep{Prusti_2016, Eyer_2018, Brown_2018, Babusiaux_2018}. 
\cite{Murphy_2019} compiled a catalogue of 1986 \textit{Kepler} $\delta$\,Scuti strs, which we added these to Fig.~\ref{fig:HR_diagram} as the grey dots. We also added the observational instability strip, defined by \cite{Murphy_2019}, based on the $\delta$\,Sct pulsator fraction, shown in Fig.~\ref{fig:HR_diagram} as purple solid lines. The black solid lines represent the theoretical instability strip of $\gamma$\,Dor stars, which was calculated by \cite{Dupret_2005} and \cite{Bouabid_2013}. Our evolutionary tracks were computed in {\sc MESA} \citep{Paxton_2011}, v10108. Our `standard model' at each mass had $X = 0.71$, $Z=0.014$ (corresponding to [M/H] $=$ 0.036), $\alpha_{\rm MLT} = 1.8$, exponential core overshooting of 0.015\,Hp, exponential H-burning shell over- and under-shooting of 0.015\,Hp, exponential envelope overshooting of 0.025\,Hp, diffusive mixing $\log D_{\rm mix} = 0$ (in cm$^{-2}$s$^{-1}$), OPAL opacities, and the \citet{Asplund_2009} solar abundance mixture. \textcolor{black}{The parameters were chosen to be close to solar values, in agreement with observed metallicity values from spectroscopy and with the values from previously published asteroseismic modelling studies of g-mode pulsators \citep[e.g.][]{Schmid_2016, Mombarg_2019}.}

Figure~\ref{fig:HR_diagram} shows that most stars (both $\delta$\,Scuti and $\gamma$\,Dor stars) are located above the zero-age main-sequence. $\delta$\,Scuti stars show more spread than $\gamma$\,Dor stars and the latter generally have lower temperatures and lower luminosities. The distribution of $\gamma$\,Dor stars is wider than theoretically predicted, especially beyond the blue edge, which shows the inconsistency to the theoretical instability strip and challenges the current stellar model. However, the conclusions are still in debate since the uncertainty of $T_\mathrm{eff}$ is $\sim250\,\mathrm{K}$. More accurate spectroscopic values for $T_\mathrm{eff}$ are required to confirm the location of the $\gamma$ Dor instability strip. The effect of binarity also needs to be considered carefully.

The $\gamma$\,Dor stars in this paper are mostly confined to the low-luminosity region near the zero-age main-sequence (ZAMS). This makes sense, because r-modes are caused by rotation, which slows down with stellar evolution after the zero-age main-sequence. Young stars rotate faster and are more likely to exhibit r-modes \textcolor{black}{in the detection region (0.2 to 2\,d)}, hence we find more r-modes in stars near the ZAMS. We do not take the possibility of pre-main sequence $\gamma$ Dor stars into account, although they were predicted by \cite{Bouabid_2011}. Another factor is that the low mass stars enter the $\gamma$ Dor instability strip at the beginning of main sequence and stay in it longer than high mass stars. Hence they are more likely to be observed. This also explains why $\gamma$ Dor stars generally show smaller $T_\mathrm{eff}$ and $\log L$.

\subsection{Period spacings}
\begin{figure*}
\centering
\includegraphics[width=\linewidth]{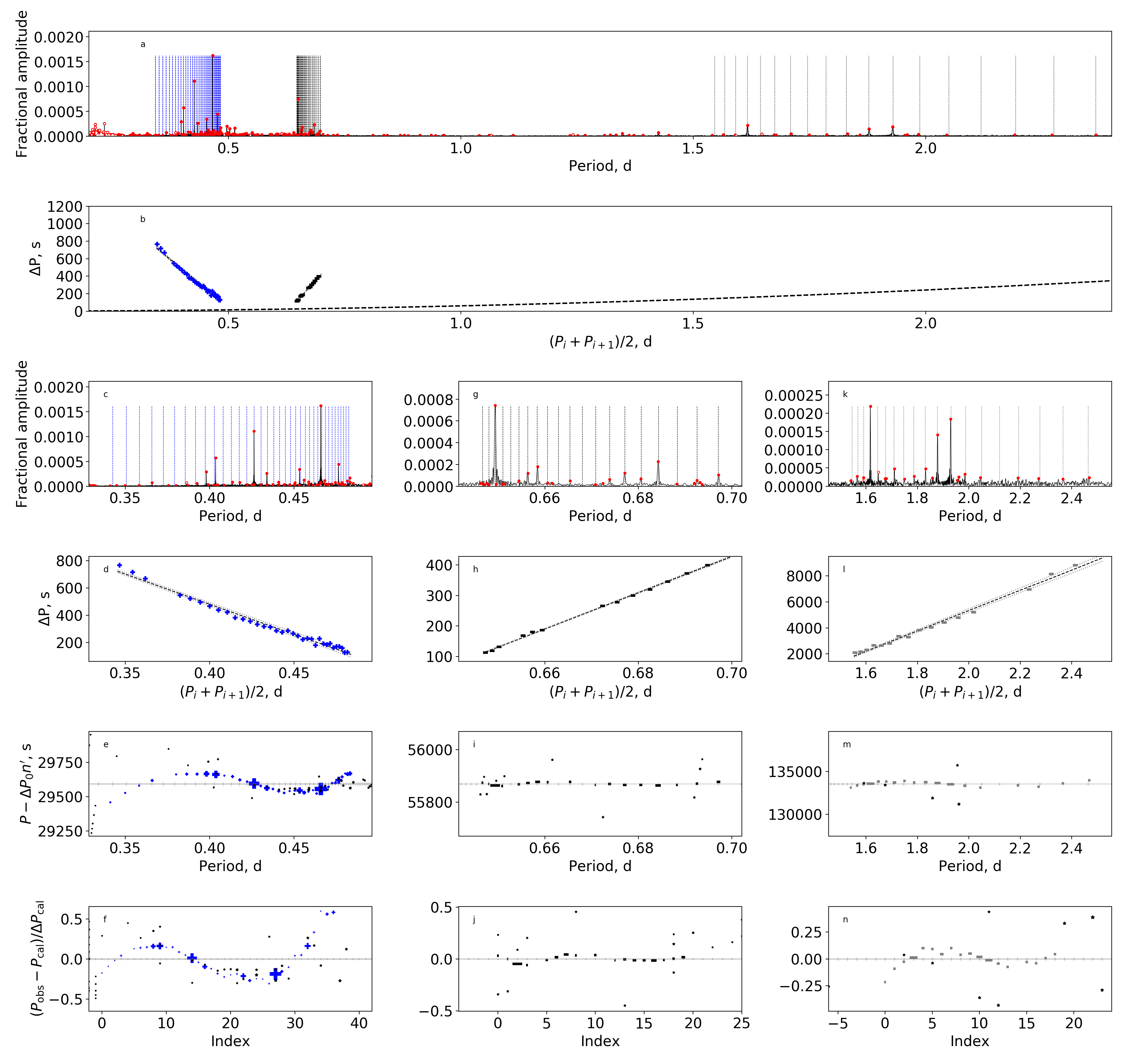}
\caption{Period spacings and the sideway \'{e}chelle diagrams of KIC 5721632. Panels (k) to (n) show the $k=-1,m=-1$ r-modes pattern. The period spacings are too larged to display in panel (b).}\label{fig:KIC5721632}
\end{figure*}

\begin{figure}
\centering
\includegraphics[width=\linewidth]{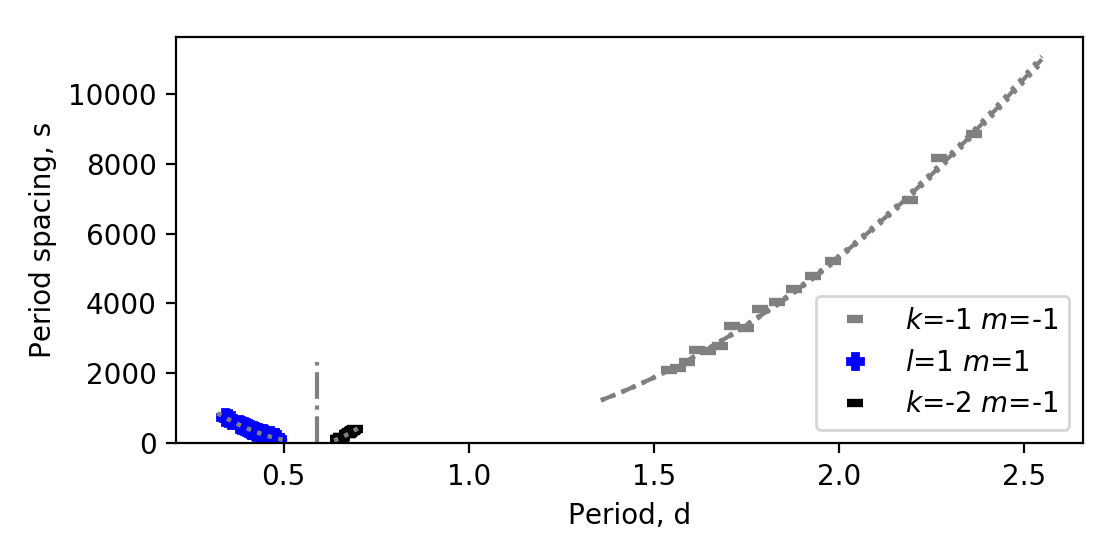}
\caption{The TAR fit of KIC\,5721632.}\label{fig:KIC5721632_tar}
\end{figure}

\begin{figure}
\centering
\includegraphics[width=\linewidth]{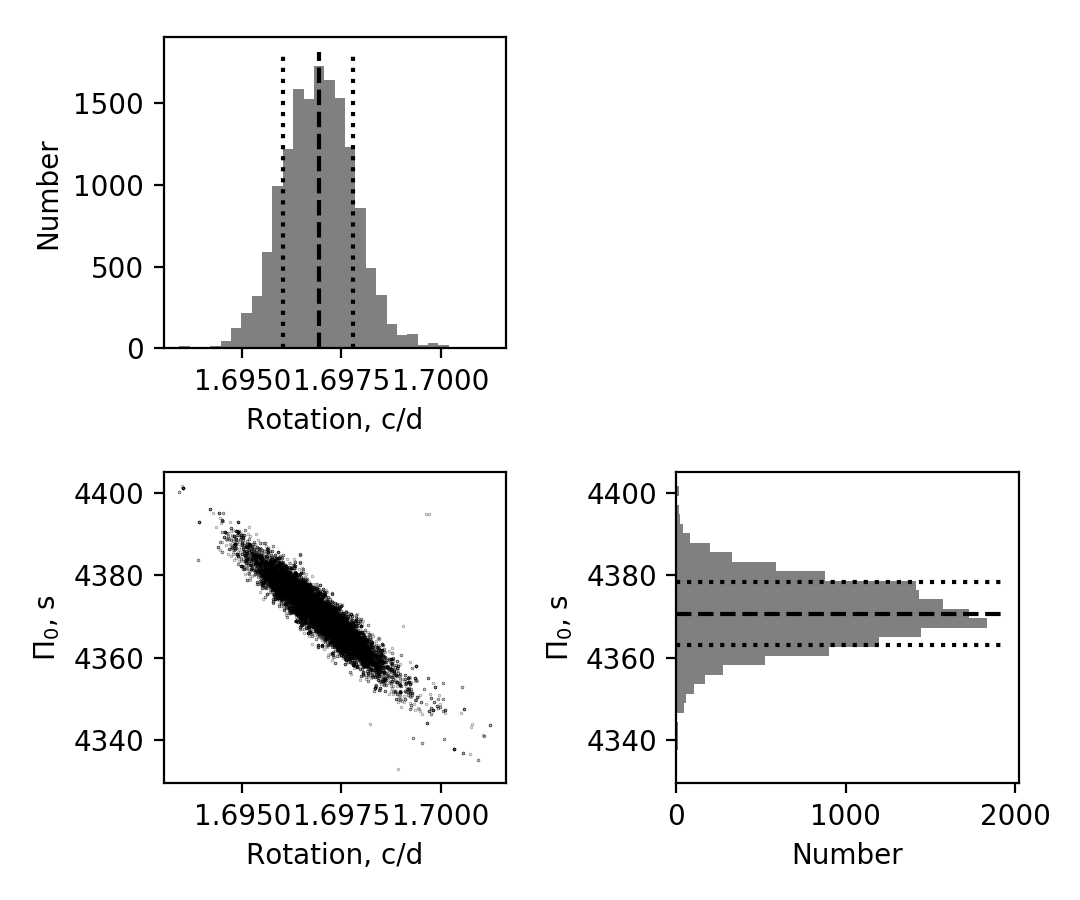}
\caption{The posterior distributions of KIC\,5721632. The best model has $\Pi_0=4371\pm8\,\mathrm{s}$ and $f_\mathrm{rot}=1.6969\pm0.0009\,\mathrm{d^{-1}}$. The fitting is dominated by the retrograde $k=-1, m=-1$ r-modes, hence $\Pi_0$ shows a negative correlation with $f_\mathrm{rot}$ in the lower-left panel.}\label{fig:KIC5721632_posterior}
\end{figure}

Our sample is large enough to show some general features for the g and r-modes of $\gamma$ Dor stars. The periodograms of r and g-mode stars show two main groups of modes. At shorter periods are the $l=1, m=1$ g-modes and at longer periods are the $k=-2, m=-1$ r-modes. Although the asymptotic spacing $\Pi_0$ of $\gamma$ Dor is typically around 4200\,s \citep{VanReeth_2016_TAR}, the period spacings vary greatly due to the rapid rotation, from a few hundred seconds to 1000\,s,  but in each star, it is similar among both the g and r-modes. The period spacings of g-modes decrease quasi-linearly with period, which is the rotational effect for prograde g-modes \citep{Bouabid_2013, Ouazzani_2017}. However, the period spacings of r-modes increase linearly with periods, and sometimes show a drop at the long-period end \citep[see e.g. Fig.~\ref{fig:KIC3240967} (h) and][]{Saio_2018_Rossby_mode}.

We found that almost all the r-modes have the quantum numbers of $k=-2, m=-1$, although other numbers were also reported by \cite{Saio_2018_Rossby_mode}. The reason is likely that the period spacing of $k=-2, m=-1$ modes is easily detected in \textit{Kepler} data, compared to tens of seconds of $k=-2, m=-2$ modes or several thousands of seconds of $k=-1, m=-1$ modes. Also, $k=-2, m=-2$ modes generally have larger amplitudes \citep{Saio_2018_Rossby_mode}. 

KIC\,5721632 is the only star which shows convincing $k=-1, m=-1$ r-modes. Figure~\ref{fig:KIC5721632} shows its amplitude periodogram, period spacing patterns and the \'{e}chelle diagram. The rotation rate is $f_\mathrm{rot}=1.6969\pm0.0009\,\mathrm{d^{-1}}$ and the asymptotic spacing is $\Pi_0=4371\pm8\,\mathrm{s}$, as shown in Figs~\ref{fig:KIC5721632_tar} and \ref{fig:KIC5721632_posterior}. In the lower-left panel of Fig.~\ref{fig:KIC5721632_posterior}, the asymptotic spacing negatively correlates with the rotation rate, since the fitting is dominated by the retrograde $k=-1, m=-1$ r-modes. The period spacing of the $k=-1, m=-1$ pattern keeps increasing from 2000\,s to 8000\,s, showing different feature from $k=-2, m=-1$ Rossby modes. 
We also noticed that there are two stars, KIC\,5721610 and KIC\,5721628, which are close to KIC\,5721632 on the sky within 6 arcseconds. We evaluated the light curve of KIC\,5721632 for contamination from those two stars. The power spectra of those two stars were calculated and we do not find any g- or r-mode oscillations. So the light curve of KIC\,5721632 was not contaminated and the $k=-1, m=-1$ Rossby modes are most likely real. 

\subsection{Asymmetric amplitude envelopes}
\cite{Saio_2018_Rossby_mode} used the visibility over the square root of kinetic energy ($\mathrm{Vis.}/\sqrt{\mathrm{K.E.}}$) to calculate the amplitude of r-modes. They showed that the $k=-2, m=-1$ modes have the highest amplitudes compared to other r-modes and the amplitude envelope is strongly asymmetric, with a steep low-period side. The relatively large amplitude is helpful to identify the $k=-2, m=-1$ r-modes. We also observed the asymmetric amplitude envelope in some stars, e.g. Fig.~\ref{fig:KIC3240967} (g), where the amplitude increases rapidly at $P\sim 0.94\,\mathrm{d}$ and drops slowly after that. However, it shows a lot of small peaks and several strong peaks before the predicted amplitude rise. For some other stars, the r-mode patterns do not show the predicted amplitude envelope, e.g. Fig.~\ref{fig:KIC5721632} (g). A more complete theory may be needed to address this observation.

\subsection{Periods of Rossby modes}
\begin{figure}
\centering
\includegraphics[width=1\linewidth]{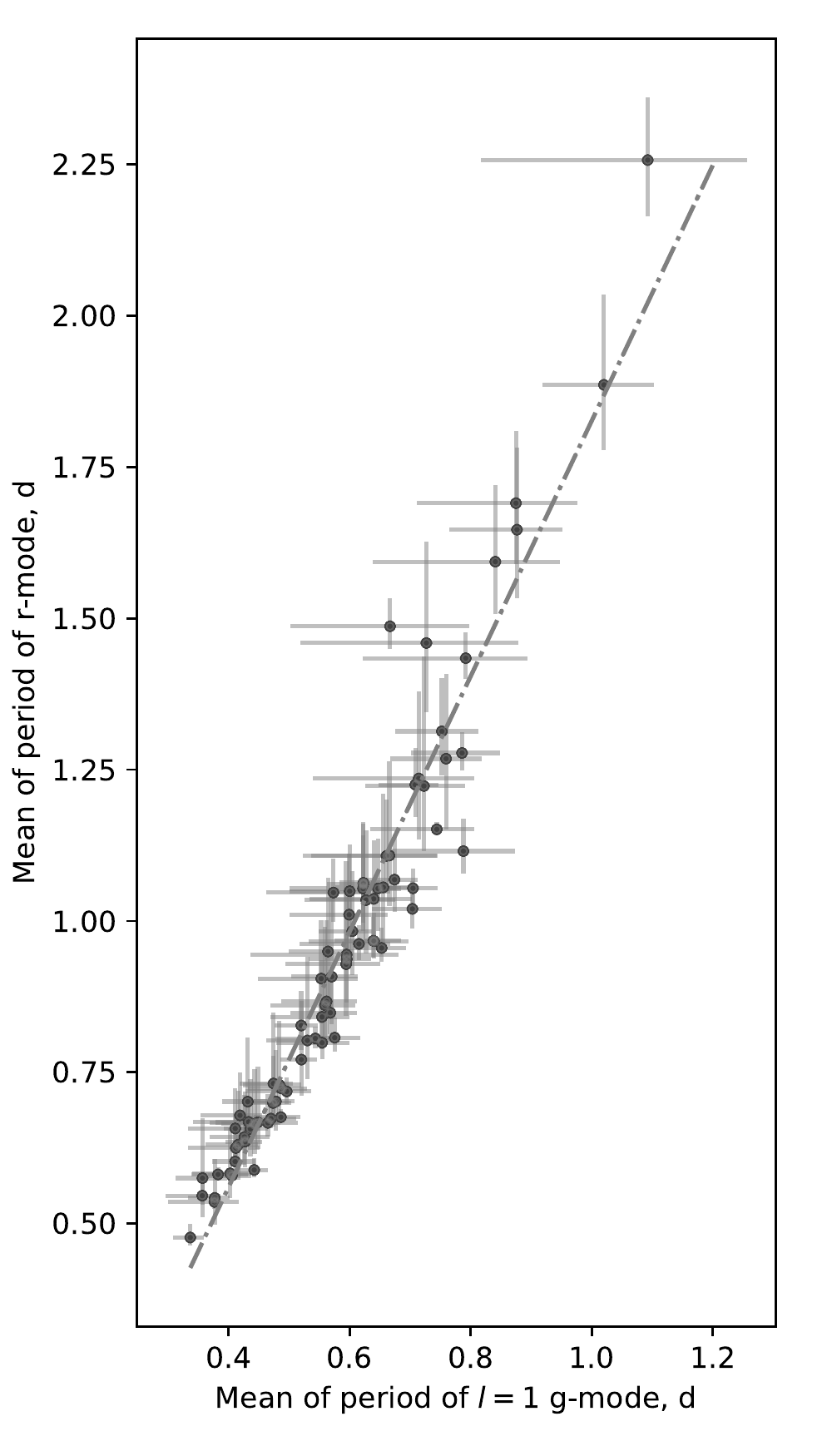}
\caption{The mean period relation between the $l=1~m=1$ g-mode patterns and the $k=-2,m=-1$ r-mode patterns. The black dots show the means of the periods in each pattern. Since the period spacing changes, the means of the periods are not in the centre. The grey solid lines are the period ranges. The dashed-dotted line is the linear fit.}\label{fig:period-period relation}
\end{figure}
Figure~\ref{fig:period-period relation} shows the period relation between the $l=1,m=1$ g-modes and $k=-2,m=-1$ r-modes. The black points are the means of the periods of each pattern, which are the averaged value of all the detected periods. Since the period spacing changes with period, the means of periods are not the centre of patterns. The grey lines are not error bars, they show the period ranges of g- and r-modes, respectively. We found that the means of the periods of g- and r-modes are related by a linear fit with 
\begin{equation}
\langle P_\mathrm{r}\rangle=\left(2.06 \pm 0.05\right) \langle P_\mathrm{g} \rangle - \left( 0.25\pm0.03 \right)\,\mathrm{d},
\end{equation}
where $\langle P_\mathrm{r} \rangle$ is the means of the periods of r-modes and $\langle P_\mathrm{g} \rangle$ is the one of g-modes. The linear fit has the slope $2.06 \pm 0.05$ and the intercept is $-0.25f\pm0.03\,\mathrm{d}$, shown as the grey dashed line. Since the means of periods must be located within the period ranges, to calculate the uncertainties of the linear fit, we used the period ranges as $3\sigma$ to make perturbations of the data. 

Based on the slope and the small intercept, we conclude the r-mode periods are close to twice the g-mode periods. The relationship helps guide a search for the r-modes in more stars, if the amplitudes of r-mode are not obvious.

\subsection{Slope--period relation}
\begin{figure*}
\centering
\includegraphics[width=0.8\linewidth]{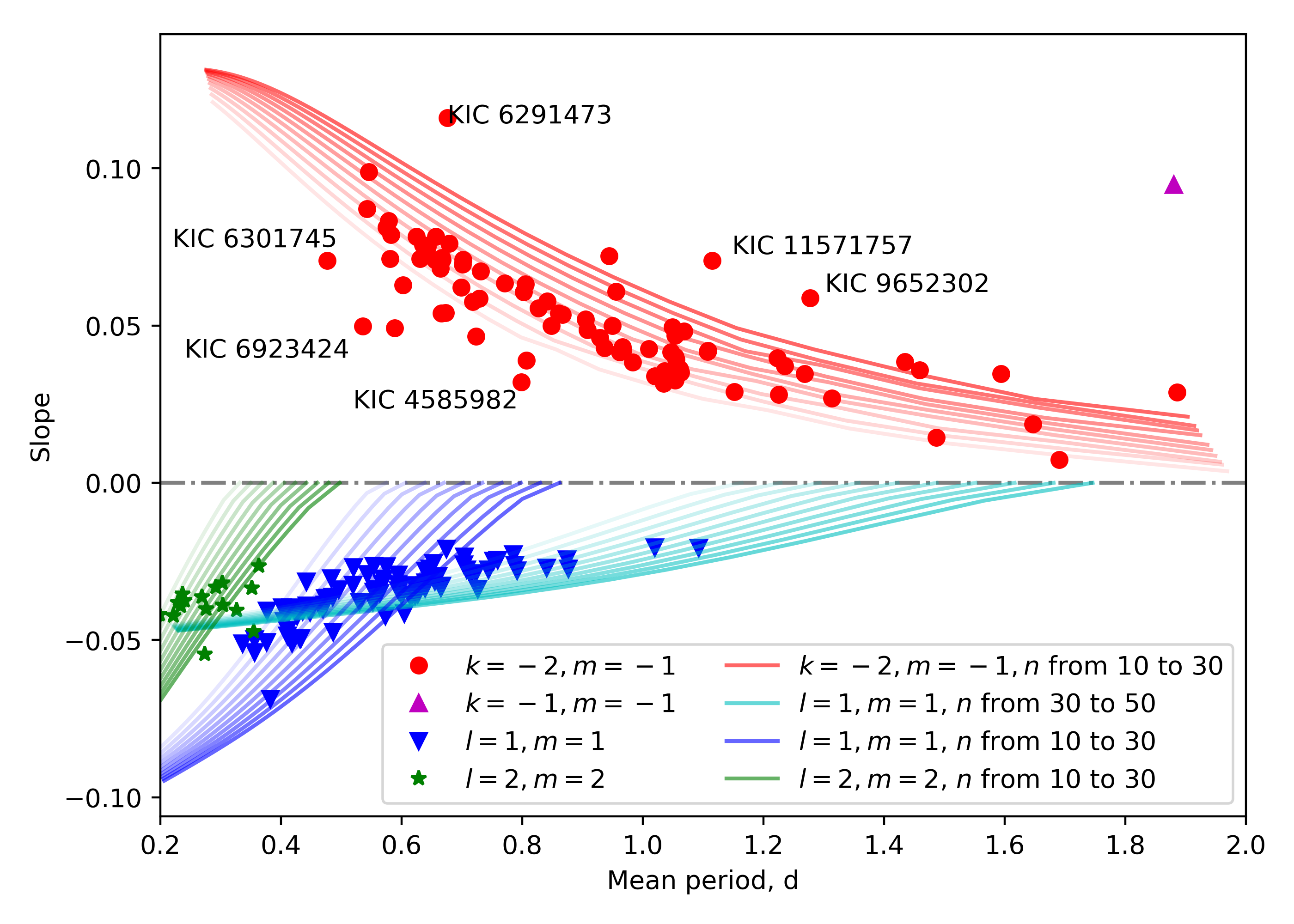}
\caption{The slope--period relation with theoretical curves for r-modes (positive slopes) and g-modes (negative slopes). We calculated the theoretical period spacing slopes in two radial order regions: 10 to 30 and 30 to 50, respectively. A given curve has the same $\Pi_0$ and $n$ region but the rotation rate increases from right to left. The transparencies of curves represent the $\Pi_0$, which is from 360\,s to 5400\,s with step of 200\,s. The lighter, the smaller the $\Pi_0$ is. Curves with different colours have different radial order ranges and different quantum numbers.}\label{fig:slope-period relation}
\end{figure*}
Figure~\ref{fig:slope-period relation} shows the slope versus period relation of \totalnumber $\gamma$ Dor stars. The x-axis is the mean of the periods, which is the same as shown in Fig.~\ref{fig:period-period relation}. The y-axis is the slope assuming the period spacing changes quasi-linearly with the period, fitted by Eq.~\ref{equ:P_i}. Different markers and colours stand for different modes. The points cluster into several groups and display different trends based on their quantum numbers. For the $l=1,m=1$ g-modes (blue triangles), the slopes are negative and increase towards zero with the mean period, with a large spread. The points of the $l=2, m=2$ g-modes (green stars) lie on the left of the dipole g-modes with similar slopes. Finally, the slopes of r-mode patterns (red circles) are larger than zero, and show a clear decreasing trend with the mean period.

To compare these results with theory, we calculated period spacing patterns from the TAR. The range of $\Pi_0$ was 3600 to 5600\,s with step of 200\,s, and $f_\mathrm{rot}$ ranged from 0 to $4\,\mathrm{d^{-1}}$. 
For r modes, we averaged over overtones with $10\leq n \leq 30$. Results are shown as red lines in Fig.~\ref{fig:slope-period relation}, with less transparency corresponding to larger $\Pi_0$. 
Considering the mean of periods and slope change with the radial order region for the g-modes, we measured the slope of calculated patterns in two regions: $10\leq n \leq 30$ and $30 \leq n \leq 50$, showing these as blue and cyan lines in Fig.~\ref{fig:slope-period relation}. Each solid line in Fig.~\ref{fig:slope-period relation} has the same $\Pi_0$ and the same $n$ region along its length, but the rotation rate increases from right to left. We use the darker lines to represent larger $\Pi_0$. The lighter the curve is, the smaller its $\Pi_0$ is. 

From the calculated curves, we find that if we want to cover the g-mode observational points, curves that span different $n$ are needed. However, the features of different radial orders differ dramatically. The blue curves with $n$ from 10 to 30 generally show smaller mean periods and steeper slopes (reaching $-0.09$ when the mean period is at $0.2\,\mathrm{d}$). The cyan curves with radial order from 30 to 50 have larger mean periods and their slopes only reach $-0.05$ when the mean period arrives at 0.2\,d. They show an overlapping area between 0.4\,d and 0.6\,d.

For the r-modes, the period spacing decreases at the long periods due to the rapid change of the eigenvalue $\lambda$. This affects the calculated slope value but is not common in the observations, so we only adopted the periods before the spacing drop, which is caused by the rapid change of the eigenvalue $\lambda$ (see Fig.~\ref{fig:KIC3240967_best_fit}). Only the results with $10\leq n \leq 30$ and $0.7\,\mathrm{d^{-1}}\leq f_\mathrm{rot} \leq 4\,\mathrm{d^{-1}}$ were used, since this parameter space covered the observations well (red lines in Fig.~\ref{fig:slope-period relation}). The radial order range from 10 to 30 of r-modes can explain the observational data. The correlation between the slope and the mean period spacing is caused by the rotation rate. The more rapidly the star rotates, the shorter the mean period and the larger the slope. In Fig.~\ref{fig:slope-period relation}, we marked six outliers, which are caused by the systematic influence of dips in the pattern. The measurement of the slope is affected strongly by the intrinsic dips or fluctuations, which reflects the limit of the linear assumption in Eq.~\ref{equ:P_i}.

\textcolor{black}{Figure~\ref{fig:slope-period relation} also shows the calculated relationships between the slope and mean period (as a function of rotation), for several discrete values of asymptotic spacing $\Pi_0$.} 
It can guide the pattern identification for other $\gamma$\,Dor stars. If a period spacing pattern does not follow the relation, several possibilities should be considered: the pattern includes a partly-observed dip \citep[e.g.][]{Li_2018}; the star is an SPB star, which generally has larger $\Pi_0$ and steeper slope \citep{Papics_2017}; the period spacing identification is misled by missing peaks.

\subsection{Asymptotic spacing and rotation}\label{subsec:Asymptotic spacing and rotation}
\begin{figure*}
\centering
\includegraphics[width=0.8\linewidth]{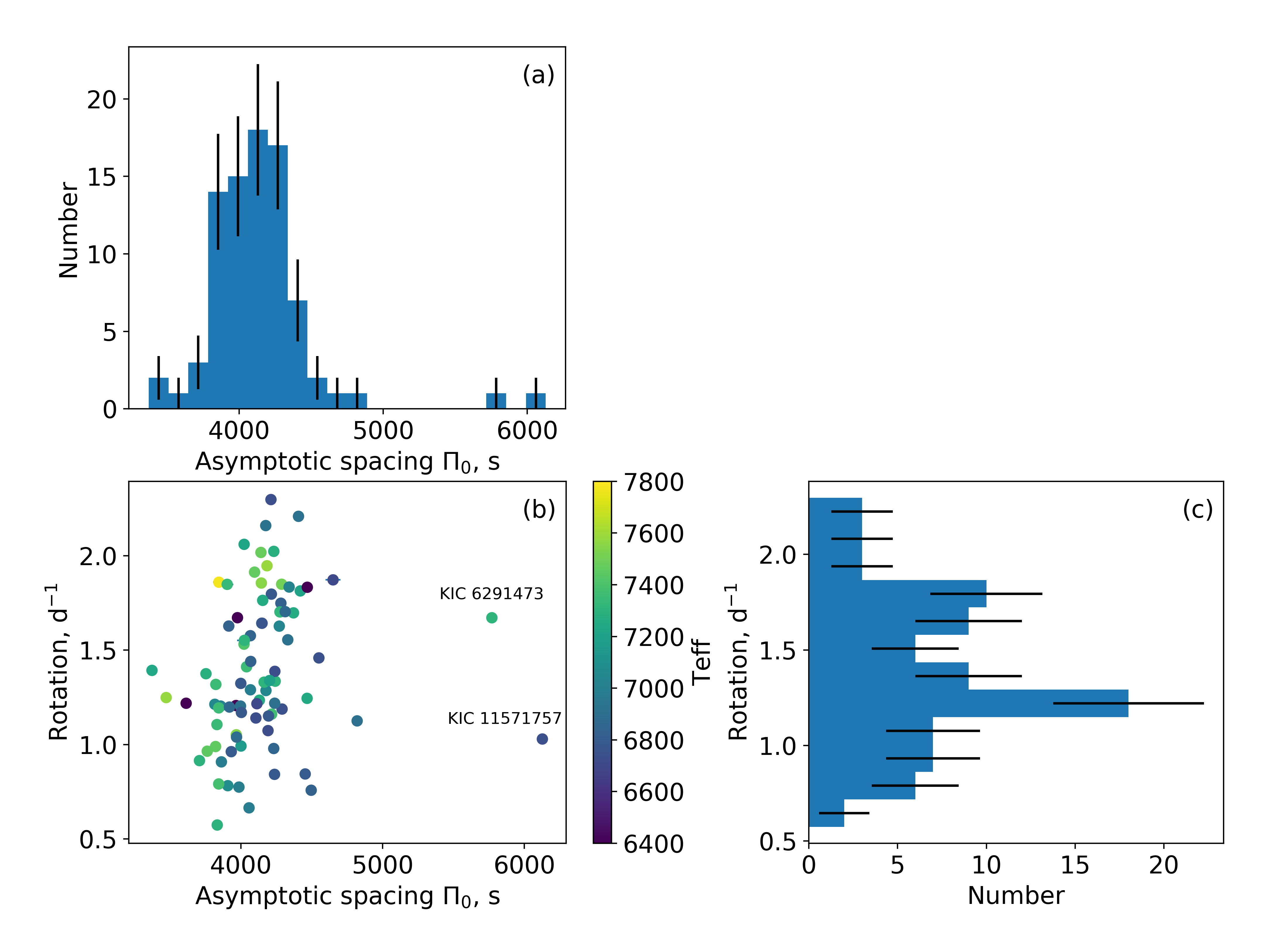}
\caption{The asymptotic spacing $\Pi_0$ and the near-core rotation $f_\mathrm{rot}$. The colour bar is the effective temperature. Uncertainties of histograms are the square roots of numbers assuming they are Poisson distributed.}\label{fig:Pi0_rot}
\end{figure*}

The TAR fit gives the near-core rotation rate $f_\mathrm{rot}$ and the asymptotic spacing $\Pi_0$ for every star. Figure~\ref{fig:Pi0_rot} shows the distributions and correlation for these two parameters. The uncertainties of the histograms are square roots of the histogram heights, assuming they are Poisson-distributed. Most stars have asymptotic spacings between 3600 and 4500\,s (Fig.~\ref{fig:Pi0_rot} (a)), which is consistent with \cite{VanReeth_2016_TAR}. The asymptotic spacing decreases with evolution, hence is considered as an indicator of stellar age. However, it is also affected by other physical parameters, such as metallicity, mixing length and convective core overshooting. Figure~\ref{fig:Pi0_rot} (c) is the distribution of the rotation rate $f_\mathrm{rot}$. The slowest rotator is KIC\,4857064 with $f_\mathrm{rot}=0.5738\pm 0.0004\,\mathrm{d^{-1}}$ and the fastest one is KIC\,6301745 with $f_\mathrm{rot}=2.297\pm 0.002\,\mathrm{d^{-1}}$. Most stars rotate between 1 and $2\,\mathrm{d^{-1}}$, much faster than the Sun but typical for A to F stars \citep{Royer_2007}.

Figure~\ref{fig:Pi0_rot} (b) displays the lack of correlation between $\Pi_0$ and $f_\mathrm{rot}$. Most stars have the asymptotic spacings around 4000\,s and the rotation rates from $1$ to $2\,\mathrm{d^{-1}}$, showing typical ranges for $\gamma$\,Dor stars. We found two outliers: KIC\,11571757, and KIC\,6291473, whose period spacing patterns are seen in the online supplementary material. The asymptotic spacings of KIC\,11571757 and KIC\,6291473 are $6130\pm30\,\mathrm{s}$ and $5770\pm40\,\mathrm{s}$, respectively. These are obviously higher than the range of $\gamma$\,Dor stars but are typical of SPB stars \citep[5600 to 15400\,s, see][]{Papics_2017}. However, their effective temperatures are too low to be B-type stars. These two stars may be at the beginning of the main sequence hence show large asymptotic spacings, or the identifications of period spacing patterns are misled by missing peaks.

We also notice that a gap at $f_\mathrm{rot}\sim 1.5\,\mathrm{d^{-1}}$ separates the data points into two groups. The $f_\mathrm{rot}$ distribution in Fig.~\ref{fig:Pi0_rot} (c) show two peaks around $1.2\,\mathrm{d^{-1}}$ and $1.8\,\mathrm{d^{-1}}$. We tried several different bins and the gap is still apparent. Whether it is a real effect or caused by the limited numbers of the sample is still an open question.

We do not find any strong correlation between $\Pi_0$ or $f_\mathrm{rot}$ and $T_\mathrm{eff}$, which is colour-coded in Fig.~\ref{fig:Pi0_rot}. Both rotation and $\Pi_0$ are expected to decrease with evolution and effective temperature, but the processes are presumably affected by many other parameters, such as the initial rotation rate, the mechanism of angular momentum transfer, or extra diffusion. It seems that \totalnumber stars are still too few to reveal a correlation. As shown in Fig.~\ref{fig:HR_diagram}, most stars in our sample are close to the ZAMS. The range of stellar ages in our sample may also be not long enough to show the correlation from the evolutionary effect. Also, more precise values of effective temperatures are still needed to confirm the theoretical prediction.

\subsection{Rotation from core to surface}
\begin{table}
\caption{The surface modulation of 6 $\gamma$\,Dor tars. `EB' means the star is a binary from \protect\cite{Kirk_2016}. `SPOT' means the surface modulation is caused by sopts. $f_\mathrm{rot, i}$ is the near-core rotation rate fitted by g- and r-mode period spacing patterns. $f_\mathrm{rot, o}$ is the surface rotation rate from spot modulation.}\label{tab:surface_rotation}
\begin{tabular}{rrrrrr}
\hline
KIC & Type & $f_\mathrm{rot, i}$, $\mathrm{d^{-1}}$   & $f_\mathrm{rot, o}$, $\mathrm{d^{-1}}$ & $f_\mathrm{rot, i}/f_\mathrm{rot, o}$ \\
\hline
KIC\,3341457 & EB & 1.859(1) & 1.893(7) & 0.982(4) \\ 
KIC\,7596250 & EB & 1.1876(7) & 1.185(4) & 1.003(4) \\ 
KIC\,7621649 & SPOT & 0.7745(4) & 0.7802(6) & 0.9928(9) \\ 
KIC\,9652302 & SPOT & 0.9147(6) & 0.910(2) & 1.005(3) \\ 
KIC\,9716563 & SPOT & 0.9081(9) & 0.90(2) & 1.01(2) \\ 
KIC\,10423501 & SPOT & 0.8420(6) & 0.841(4) & 1.001(4) \\ 
\hline
\end{tabular}
\end{table}

\begin{figure}
\centering
\includegraphics[width=\linewidth]{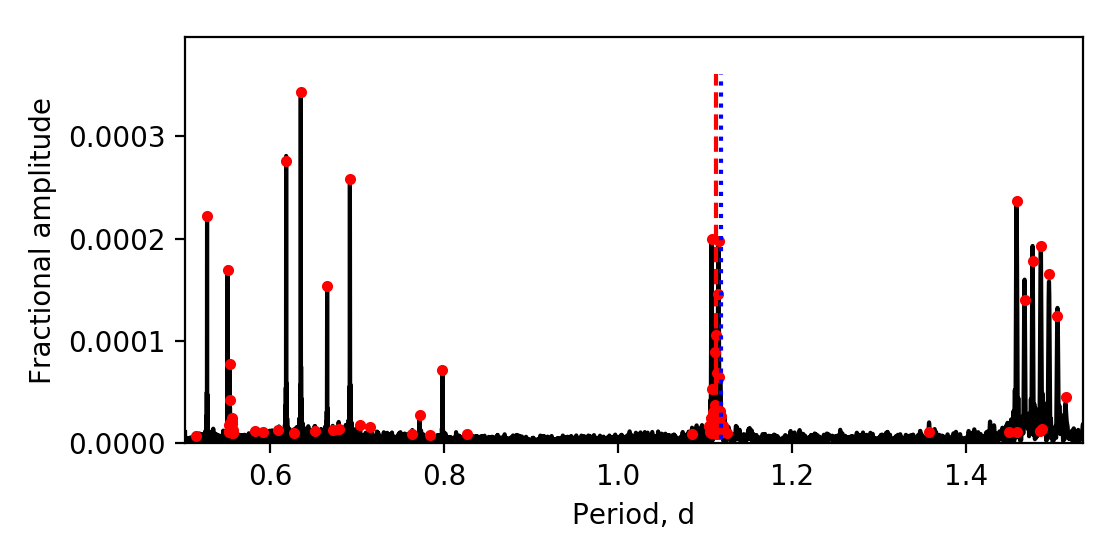}
\caption{The pulsation and surface rotation of KIC\,9716563. Three peak groups are seen. The peak groups at 1.11\,d is the signal of the surface modulation. The blue dotted line (overlapped with the red dotted line) displays the near-core rotation while the red dashed line shows the surface rotation, which is the mean of the peaks. The periods below 0.8\,d are $l=1,m=1$ g-modes and the periods above 1.4\,d are $k=-2, m=-1$ r-modes.}\label{fig:KIC9716563_surface_rotation}
\end{figure}

\begin{figure}
\centering
\includegraphics[width=\linewidth]{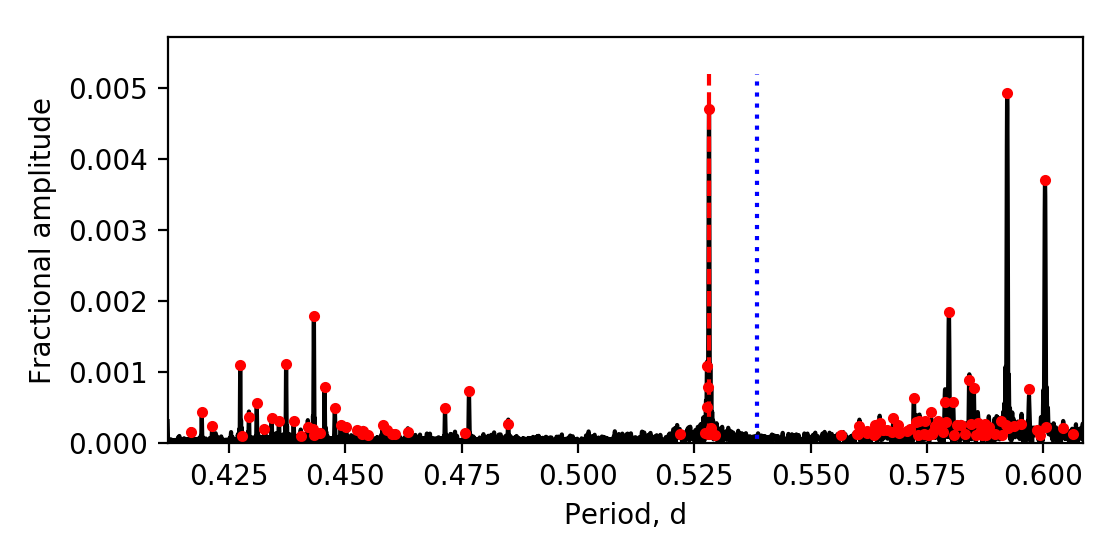}
\caption{The pulsation and surface rotation of KIC\,3341457. This star shows a large discrepancy between the near-core rotation period (blue dotted line) and the orbital period (red dashed line). It implies a radial differential rotation. }\label{fig:KIC3341457_surface_rotation}
\end{figure}

The rotation profile from the core to the surface is the key to understand the angular momentum transport and chemical mixing processes. As pointed out in Section~\ref{subsec:Asymptotic spacing and rotation}, the g- and r-mode period spacings are used to constrain near-core rotation rates inside the stars. Several methods give the opportunity to measure the rotation of the outer stellar envelope, such as 
p-mode splittings \citep[e.g.][]{Kurtz_2014, Saio_2015, Schmid_2016}, 
projected rotation velocity ($v \sin i$) \citep[e.g.][]{Murphy_2016}, 
rotational spot modulation \citep[e.g.][]{Degroote_2011, McQuillan_2014, Garcia_2014, Balona_2017}, 
and deducing surface rotation by tidal locking in binaries \citep[e.g.][]{Kallinger_2017, Guo2017, Zhang_2018}. 
A signal caused by spots is common for low-mass stars with convective envelopes. However, only 10\% of intermediate- and high-mass stars show detectable magnetic fields based on Zeeman splitting \citep{Donati_2009, Wade_2016}, which implies a lack of spot activity. Nevertheless, \cite{VanReeth_2018} detected rotational modulations from 8 stars among 68 $\gamma$\,Dor stars, whose detection percentage is consistent with the magnetic observation.

The signal of rotational modulation in frequency is located between g- and r-modes and we have searched for it in all stars. We follow the criteria reported by \cite{VanReeth_2018} to inspect the surface rotation. The criteria are:
(1) the S/N of the peak is larger than 4; 
(2) the second harmonic $2f$ is found; 
(3) the selected frequency is part of an isolated group of frequencies, since the lifetime of spot is assumed to be shorter than the Kepler data span; 
(4) the peak should stand between g- and r-modes. 
We found six stars which have a signal of rotational modulation that satisfies all these conditions. Table~\ref{tab:surface_rotation} lists the near-core and surface rotation rates ($f_\mathrm{rot, i}$ and $f_\mathrm{rot, o}$) and their ratio. KIC\,3341457 and KIC\,7596250 are classified as eclipsing binaries or ellipsoidal variables \citep{Kirk_2016} while the modulations for the other three stars are likely caused by spots. Only KIC\,3341457 has an obvious differential rotation. The other four stars are quasi-rigidly rotating, as their derived $f_\mathrm{rot, i}$ and $f_\mathrm{rot, o}$ values are close to each other. 

Appendix~\ref{appendix: surface} contains their periodograms. We use KIC\,9716563 as an example. Figure~\ref{fig:KIC9716563_surface_rotation} displays the periodogram of KIC\,9716563. The peaks at periods below 0.8\,d belong to the $l = 1, m = 1$ g-modes and the peaks at periods above 1.4\,d are $k = −2, m = −1$ r-modes. There is a region between g- and r-modes in which we did not find any extra peaks. However, a strong peak group appears at 1.11\,d, which is roughly equal to the near-core rotation rate $1.102 \pm 0.001\,\mathrm{d}$. We identified this peak group as the surface modulation, which is likely to be caused by surface spot. The surface modulation shows multiple peaks, indicating the spot lifetime is shorter than the data span, the spot size varies with time, or there is latitude differential rotation. The amplitude caused by the spot is similar to the strongest g and r-mode oscillations ($\sim0.2\%$ of the mean flux).

KIC\,3341457 is the only star which shows differential rotation. It is identified as `an ellipsoidal variable or a system with an uncertain classification' with orbital period of 0.5281792\,d \citep{Matijevic_2012, Kirk_2016}.
Figure~\ref{fig:KIC3341457_surface_rotation} displays the periodogram of KIC\,3341457. The orbital effect generates a peak group at 0.528\,d. However, the near-core rotation derived from the g- and r-modes is 0.538\,d, 2\% longer than the surface rotation. Considering the short orbital period, the stellar surface is likely tidally locked hence this star has a non-uniform rotational profile. 

KIC\,7596250 is also identified as a binary but its orbital period is equal to the near-core rotation period. So KIC\,7596250 is rigidly rotating, assuming its surface is tidally locked.

Although spot activity is generally weak in intermediate-mass stars, it is still possible to find surface modulations in a few cases. Apart from the spot activity, binary effect also involves a signal corresponding to the orbital period which implies surface rotation, assuming the surface is tidally locked. According to our results and those from \cite{VanReeth_2018}, most main-sequence stars have uniform rotation profiles, while the components in binary systems are more likely to be differentially rotating, due to the tidal locking effect. 
\textcolor{black}{Many theoretical mechanisms can cause the differential rotation, such as the internal gravity waves \citep{Aerts_2018}, magnetic fields \citep[e.g.][]{Mathis_2005, Prat_2019}, and tides \citep[e.g.][]{Lurie_2017, Zhang_2018}. KIC\,3341457 is an excellent star for modelling angular momentum transport mechanisms. }

\subsection{Radial order distribution}

\begin{figure}
\includegraphics[width=1\linewidth]{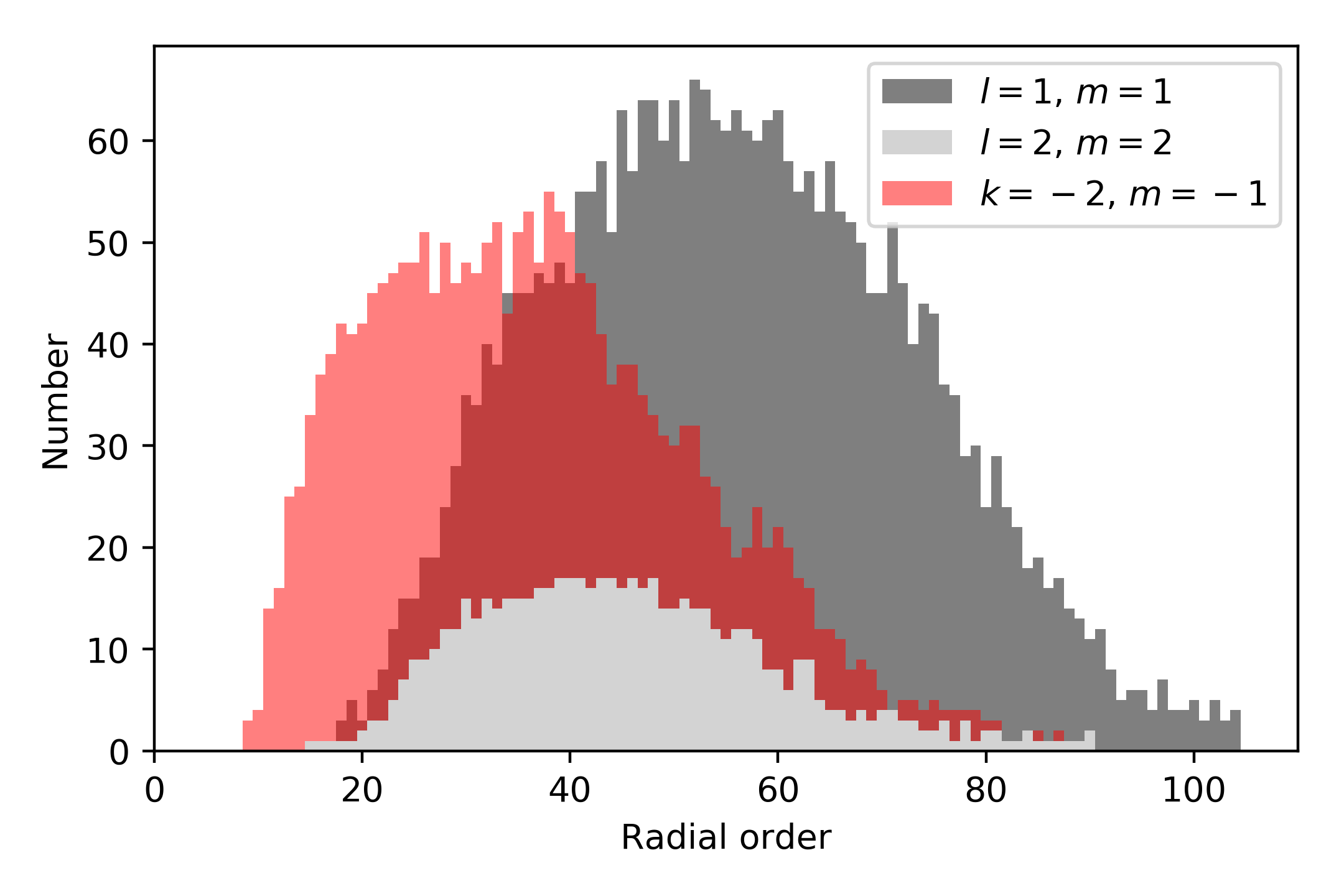}
\caption{The distributions of radial orders. The black, white and red areas stand for dipole, quadrupole g-modes and r-modes.}\label{fig:radial_order_distribution}
\end{figure}

\begin{figure}
\includegraphics[width=\linewidth]{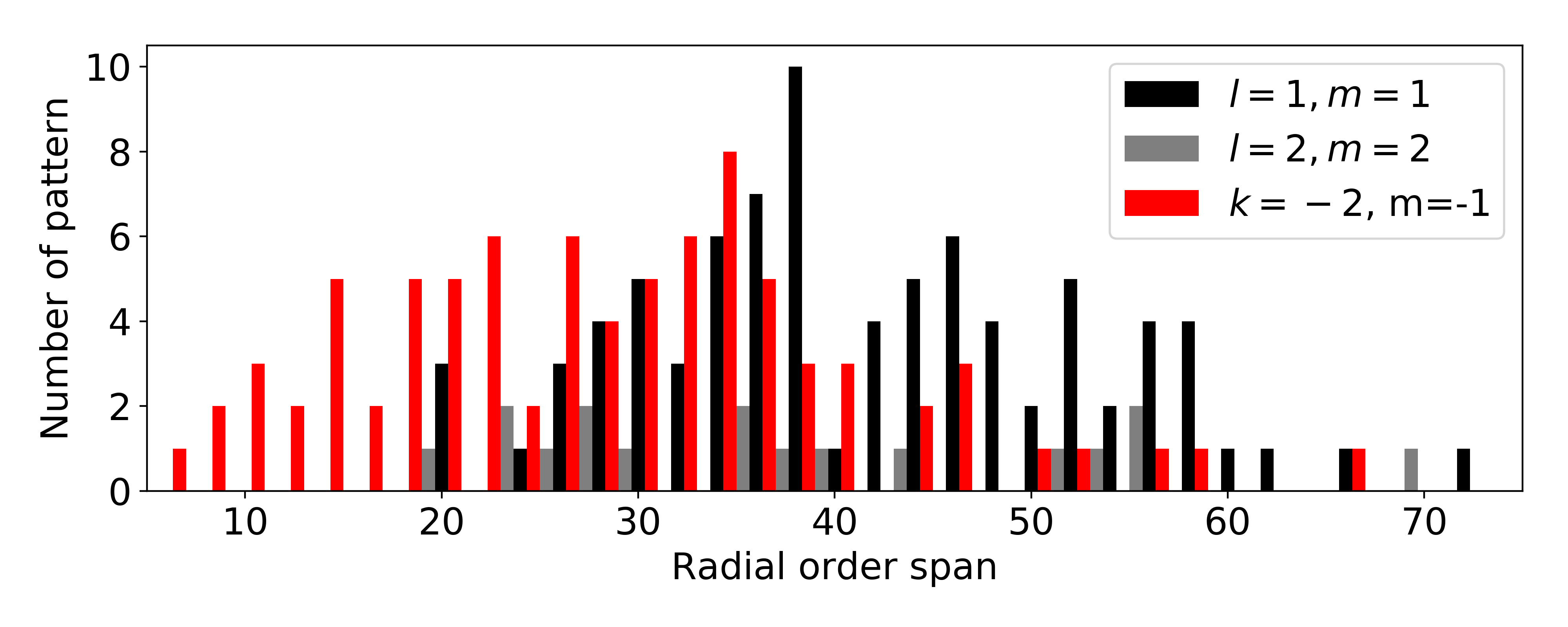}
\caption{The radial order span of the g- and r-modes. Most of patterns show $\sim30$ radial order length. But several g-modes patterns show extremely long radial order spans, even up to 70.}\label{fig:pattern_length}
\end{figure}

The radial orders of pulsations in a slow rotator can be estimated as the period divided by the period spacing, $P/\Delta P$, because of the asymptotic spacing in Eq.~\ref{equ:asymptotic spacing}. However, rotation changes the period spacing, hence this becomes invalid for most $\gamma$\,Dor stars. We used the TAR fit results to get more accurate estimates of the radial orders. Fig.~\ref{fig:radial_order_distribution} displays the radial order distributions of the excited modes for the whole sample.

For dipole g-modes, the most likely radial order is $\sim55$. The distribution shows a spread with Full-Width-at-Half-Maximum (FWHM) of $\sim50$. The lowest $n$ is 18 while the highest one is 103. 
For quadrupole g-modes, the range in radial orders spans from 15 to 83 with FWHM of $\sim30$. The most frequent radial order is $\sim40$, slightly lower than for dipole g-modes, which is possibly because identification of quadrupole modes is more difficult, since the period region overlaps with the harmonics of dipole g modes, and the peaks are generally crowded. This may lead to missing the highest order modes. 

Both the dipole and quadrupole g modes are symmetrically distributed. However, the distribution of radial orders of r-modes is asymmetric. Most r-modes peaks have the radial order between $\sim13$ and $\sim53$ with FWHM of $\sim40$, which is lower than the g modes.

Figure~\ref{fig:pattern_length} summarises the pattern lengths of the dipole, quadrupole g-modes and r-modes, determined as the difference between the largest and smallest radial orders in each star. For the dipole g modes (black histogram), patterns typically span $\sim35$. Of those, KIC\,11571757 has the longest period spacing pattern spanning 73 modes, as shown in the online supplementary material. For the quadrupole g-mode patterns (grey), the typical length is similar to the dipole one. For the r-modes (red), the pattern is typically shorter than for g-modes. The distribution of r-modes shows the excess over that of g-modes when the length is smaller than 30. Of those, KIC\,5640438 shows the longest r-mode period spacing pattern, which spans 67 radial orders.

To summarise, the distributions of excited radial orders in g-modes is symmetric with centre of $n\sim40$. There are more low-radial-order r-modes than g-modes, and their distribution is asymmetric. The excited modes generally span tens of radial orders in both g- and r-modes. Rossby mode patterns show lower radial orders and shorter lengths than the g-mode patterns. 

\section{conclusions}\label{sec:conclusions}
We report \totalnumber $\gamma$\,Dor stars in which the period spacing patterns of both g- and r-modes are seen, which forms the
largest sample of $\gamma$\,Dor stars with r-modes. The period spacing changes approximately linearly with period due to the near-core rotation but shows some fluctuations caused by chemical composition gradients. In each star, the g- and
r-modes have similar period spacings (around several hundred seconds), but the mean periods of r-modes are generally twice those of g-modes. We observed some turning points in $\Delta P$ for r-modes in several stars, caused by the rapid change of the eigenvalue $\lambda$. By fitting a straight line to $\Delta P$ versus $P$, we find that the slope correlates negatively with the mean period in r-modes. 

The traditional approximation of rotation (TAR) was used to fit the observed patterns and calculate the near-core rotation rates. The fit matched the observed patterns well, and the near-core rotation rate ($f_\mathrm{rot}$) and the asymptotic spacing ($\Pi_0$) were constrained accurately. Most of the \totalnumber stars have rotation frequencies between $1$ and $2\,\mathrm{d^{-1}}$, with $\Pi_0$ around 4000\,s. We noticed six stars which have surface modulations caused by spots or the ellipsoidal variability in a binary system. Of those, five stars rotate uniformly while KIC\,3341457 has a differential core-to-surface rotation. This is consistent with the previous observations that most main-sequence stars rotate quasi-rigidly while components in binary systems are more likely to have non-uniform rotation profile. The radial orders of excited modes display different distributions for the dipole g-modes, quadrupole g-modes, and r-modes. Most g-modes appear around $n \sim 55$, while the radial orders of r-modes are likely to be between 13 and 50, generally lower than those of g-modes.

Our sample is very useful for understanding the physics of rotation and angular momentum transport in main sequence stars, and can place much needed constraints on diffusive mixing and chemical gradients, whilst also providing stellar ages.

\section*{Acknowledgement}
The research was supported by an Australian Research Council DP grant DP150104667. Funding for the Stellar Astrophysics Centre is provided by the Danish National Research Foundation (Grant DNRF106). This work has made use of data from the European Space Agency (ESA) mission
{\it Gaia} (\url{https://www.cosmos.esa.int/gaia}), processed by the {\it Gaia}
Data Processing and Analysis Consortium (DPAC,
\url{https://www.cosmos.esa.int/web/gaia/dpac/consortium}). Funding for the DPAC
has been provided by national institutions, in particular the institutions
participating in the {\it Gaia} Multilateral Agreement.

\bibliographystyle{mnras}
\bibliography{ligangref} 

\appendix
\section{Period spacing diagrams}\label{appendix: period spacing diagrams}
We show the period spacing patterns, the sideways \'{e}chelle diagrams, the TAR fit results, and the posterior distributions of \totalnumber $\gamma$\,Dor stars with gravity modes and Rossby modes patterns. Example diagrams are shown in figs A1 to A3. The figures for the other \totalnumberminusone stars are available as supplementary online material.
\begin{figure*}
\centering
\includegraphics[width=1\linewidth]{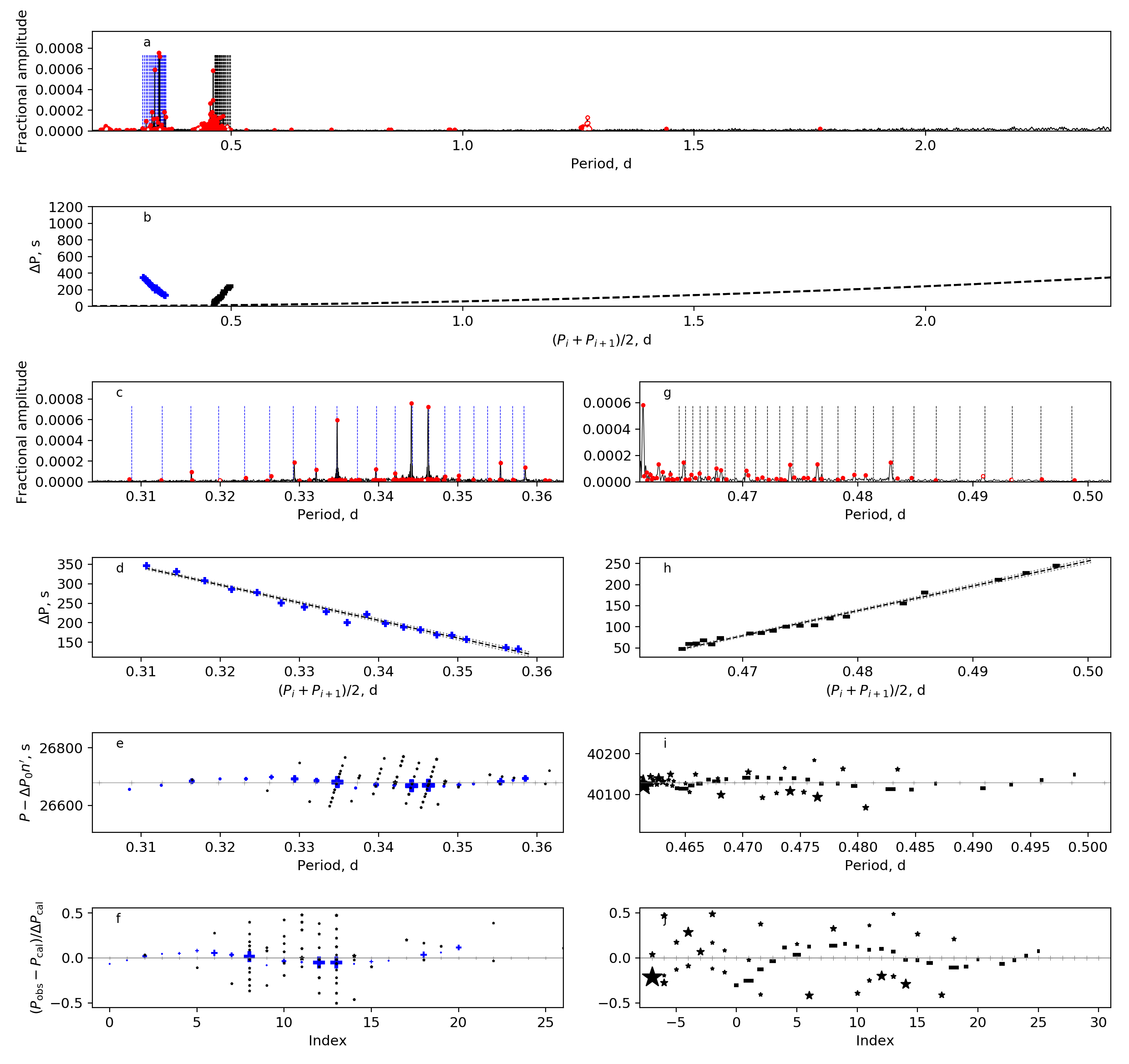}
\caption{The g and r-mode patterns of KIC\,6301745. Panel a: the amplitude spectrum with x-axis of period. The solid red circles present the detected independent frequencies while the open red circles show the combination frequencies. The vertical dashed lines are the linear fits for each pattern. The x-axis range is set from 0.2\,d to 2.4\,d for consistency for all stars. There are two period spacing patterns. The blue one on the left is the $l=1, m=1$ g-modes while the black one on the right is the $k=-2, m=-1$ r-modes. Panel b: the period spacing patterns of KIC\,6301745. The linear fits and uncertainties are shown by the black and grey dashed lines. The blue plus symbols are the g-modes and the black minus symbols are the r-modes. The dashed line is the period resolution. Panels c and d: the detail of the spectrum and period spacing pattern of g-modes. Panel e: the sideways \'{e}chelle diagram of the g pattern. Panel f: the normalised sideways \'{e}chelle diagram of the g-modes pattern. Panels g to j: same but for the r-mode pattern.  }\label{appfig:KIC6301745}
\end{figure*}

\begin{figure}
\centering
\includegraphics[width=\linewidth]{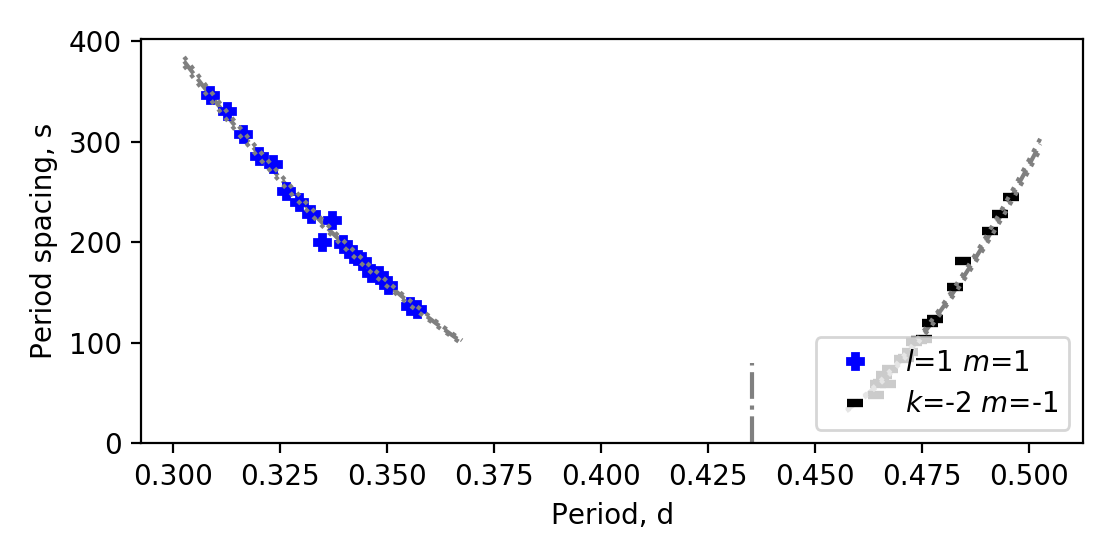}
\caption{The TAR fit of KIC\,6301745. The blue plus symbols show the $l=1, m=1$ gravity modes and the black minus symbols show the $k=-2, m=-1$ Rossby modes. The dashed grey curves display the best-fitted result. The error margins are plotted by the dotted lines. The vertical dashed line denotes the fitted rotation period. }\label{appfig:KIC6301745_best_fit}
\end{figure}

\begin{figure}
\centering
\includegraphics[width=\linewidth]{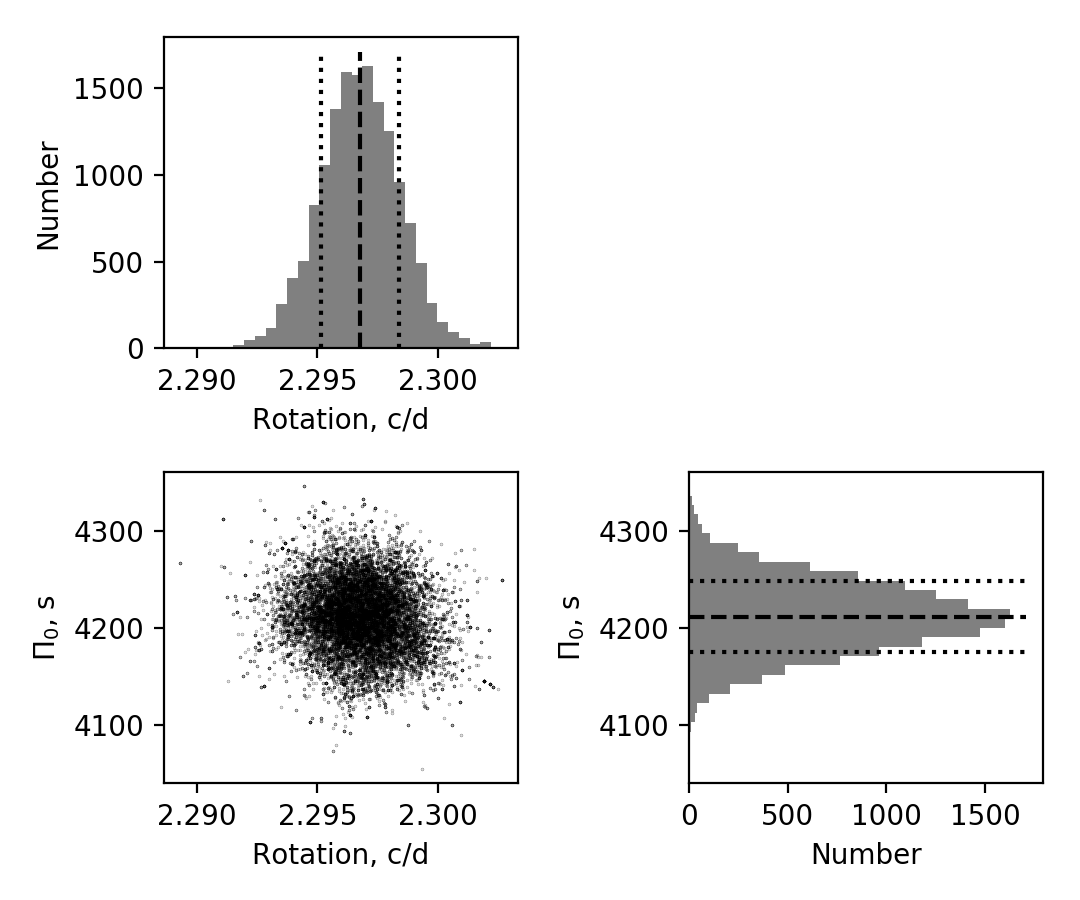}
\caption{The posterior distributions for the TAR fit to KIC\,6301745 using eq.~\ref{equ:likelihood}. The dashed lines are the medians and the dotted lines show the $\pm 1\sigma$ areas. }\label{appfig:KIC6301745_posterior}
\end{figure}

\section{surface modulations}\label{appendix: surface}

We present six stars with surface modulations in this appendix. KIC\,3341457 displays a strong radial differential
rotation while the other five stars show uniform rotation.

\begin{figure}
\centering
\includegraphics[width=\linewidth]{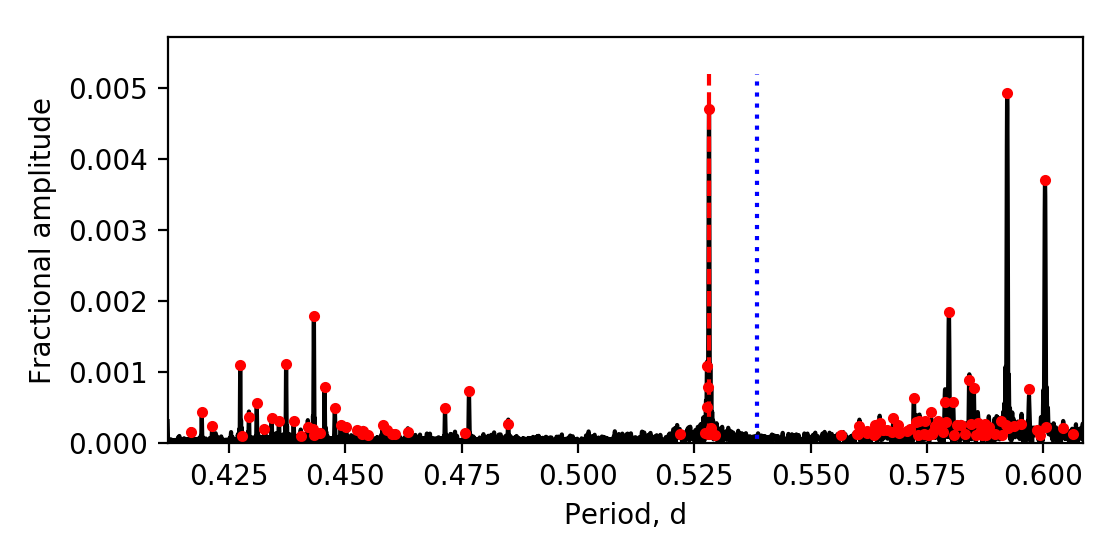}
\caption{The periodogram and surface rotation of KIC\,3341457. The red dots are peaks with S/N>4. The peak groups at 0.528\,d is the signal of the surface modulation. The blue dotted line displays the near-core rotation while the red dashed line shows the surface rotation, which is the mean of the peaks. }\label{appfig:KIC3341457_surface_rotation}
\end{figure}

\begin{figure}
\centering
\includegraphics[width=\linewidth]{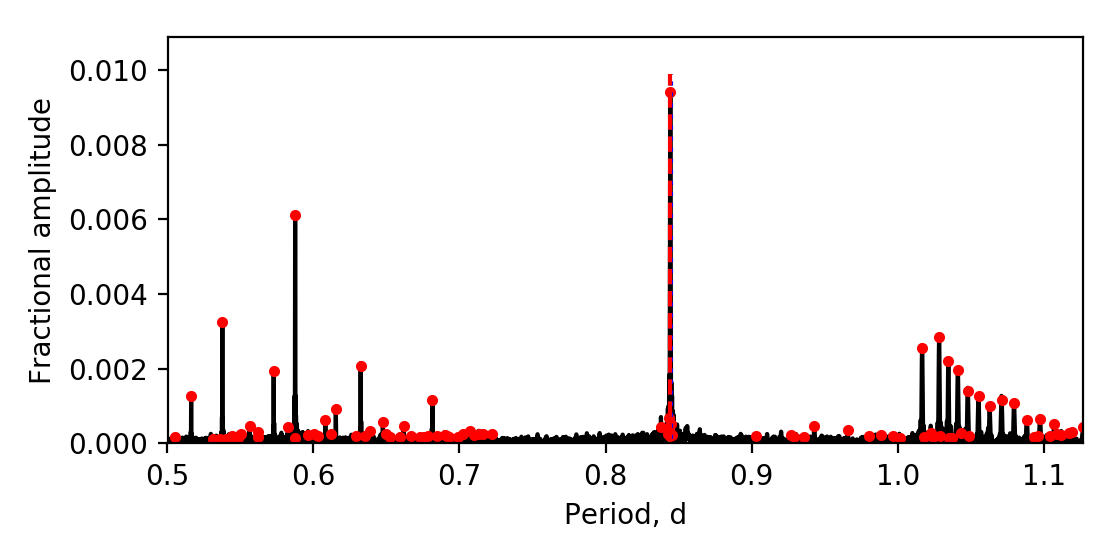}
\caption{The periodogram and surface rotation of KIC\,7596250. }\label{appfig:KIC7596250_surface_rotation}
\end{figure}

\begin{figure}
\centering
\includegraphics[width=\linewidth]{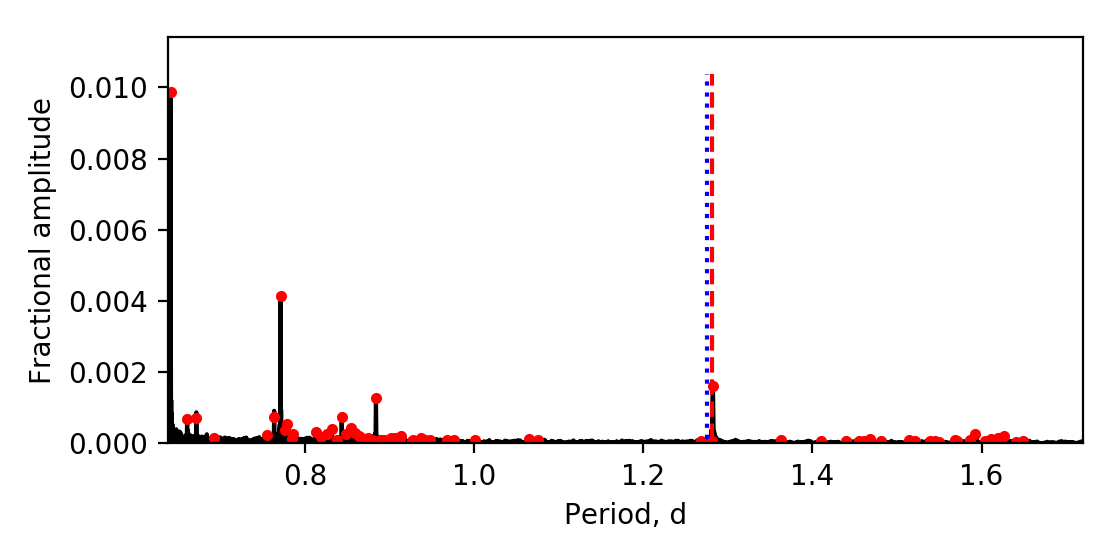}
\caption{The periodogram and surface rotation of KIC\,7621649. }\label{appfig:KIC7621649_surface_rotation}
\end{figure}

\begin{figure}
\centering
\includegraphics[width=\linewidth]{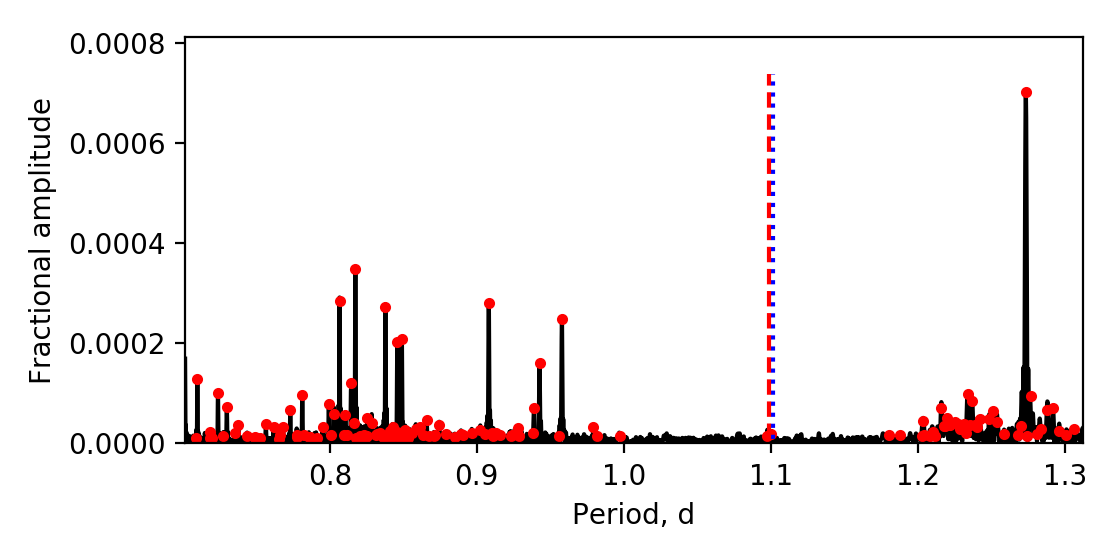}
\caption{The periodogram and surface rotation of KIC\,9652302. }\label{appfig:KIC9652302_surface_rotation}
\end{figure}

\begin{figure}
\centering
\includegraphics[width=\linewidth]{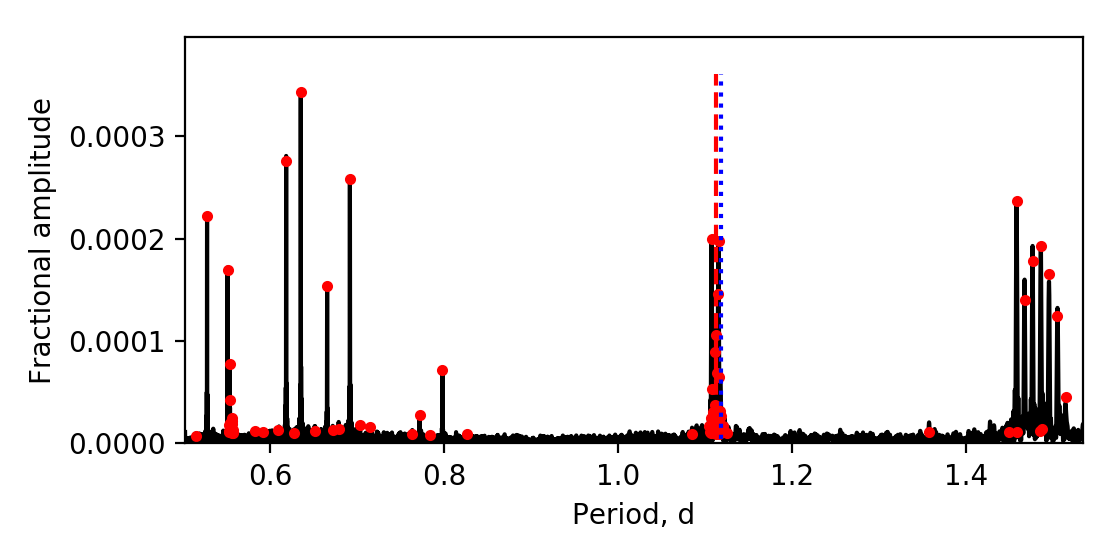}
\caption{The periodogram and surface rotation of KIC\,9716563. }\label{appfig:KIC9716563_surface_rotation}
\end{figure}

\begin{figure}
\centering
\includegraphics[width=\linewidth]{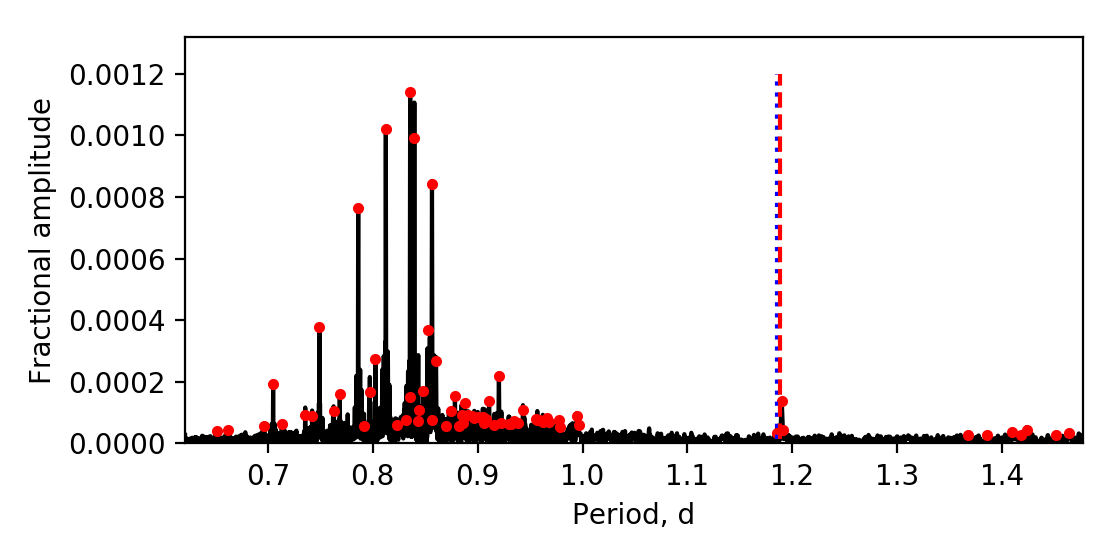}
\caption{The periodogram and surface rotation of KIC\,10423501. }\label{appfig:KIC10423501_surface_rotation}
\end{figure}

\bsp	
\label{lastpage}
\end{document}